\documentclass[
times,
aps,
twocolumn,
showpacs,
superscriptaddress,
prx,
floatfix]{revtex4-2}

\usepackage{graphicx}
\usepackage{array}
\usepackage{mathtools}
\usepackage{color}
\usepackage{upgreek}
\usepackage{float}
\usepackage{enumerate}
\usepackage{enumitem}
\usepackage{lipsum}
\usepackage{hhline}
\usepackage{booktabs}
\usepackage{url}
\usepackage{braket}

\usepackage{amsthm}
\usepackage{bm}
\usepackage[T1]{fontenc}
\usepackage{scrextend}
\usepackage{hyperref}
\usepackage[dvipsnames]{xcolor}
\definecolor{C1}{RGB}{52, 89, 149}
\definecolor{C2}{RGB}{251, 77, 61}
\definecolor{C3}{RGB}{3, 206, 164}
\definecolor{C4}{RGB}{202, 21, 81}
\hypersetup{colorlinks=true, linkcolor=C2, citecolor=C2, urlcolor=C2}
\usepackage{dsfont}
\usepackage{epstopdf}
\usepackage{tikz}
\usepackage{mathrsfs}
\usepackage[caption=false]{subfig}
\usepackage{fdsymbol}

\usepackage{accents}

\theoremstyle{remark}

\newcommand*{\ups}{\Upsilon}

\newcommand*{\id}{\mathds{1}}

\DeclareMathOperator{\tr}{tr}

\begin{document}
\title{Diagnosing chaos with projected ensembles of process tensors}

\author{Peter O'Donovan}
\email{peodonov@tcd.ie}
\affiliation{School of Physics, Trinity College Dublin, College Green, Dublin 2, D02 K8N4, Ireland}

\author{Neil Dowling}
\email[]{ndowling@uni-koeln.de}
\affiliation{Institut f\"ur Theoretische Physik, Universit\"at zu K\"oln, Z\"ulpicher Strasse 77, 50937 K\"oln, Germany}
\affiliation{School of Physics \& Astronomy, Monash University, Clayton, VIC 3800, Australia}

\author{Kavan Modi}
\email{kavan@quantumlah.org}
\affiliation{Science, Mathematics and Technology Cluster, Singapore University of Technology and Design, \\8 Somapah Road, 487372 Singapore}
\affiliation{School of Physics \& Astronomy, Monash University, Clayton, VIC 3800, Australia}

\author{Mark T. Mitchison}
\email{mark.mitchison@kcl.ac.uk}
\affiliation{School of Physics, Trinity College Dublin, College Green, Dublin 2, D02 K8N4, Ireland}
\affiliation{Department of Physics, King’s College London, Strand, London, WC2R 2LS, United Kingdom}

\date{\today}

\begin{abstract}
The process tensor provides a general representation of a quantum system evolving under repeated interventions and is fundamental for numerical simulations of local many-body dynamics. In this work, we introduce the projected process ensemble, an ensemble of pure output states of a process tensor in a given basis of local interventions, and use it to define increasingly more fine-grained probes of quantum chaos. The first moment of this ensemble encapsulates numerous previously studied chaos quantifiers, including the Alicki-Fannes quantum dynamical entropy, butterfly flutter fidelity, and spatiotemporal entanglement. We discover characteristic entanglement structures within the ensemble's higher moments that can sharply distinguish chaotic from integrable dynamics, overcoming deficiencies of the quantum dynamical and spatiotemporal entropies. These conclusions are supported by extensive numerical simulations of many-body dynamics for a range of spin-chain models, including non-interacting, interacting-integrable, chaotic, and many-body localized regimes. Our work elucidates the fingerprints of chaos on spatiotemporal correlations in quantum stochastic processes, and provides a unified framework for analyzing the complexity of unitary and monitored many-body dynamics. 

\end{abstract}

\maketitle

\section{Introduction}

Quantum chaos has become synonymous with randomness. Probes such as level spacing statistics~\cite{bgs1984, borgonovi_quantum_2016}, eigenstate properties~\cite{ srednicki_chaos_1994, Rigol2016, Vidmar2017, Bianchi_Hackl_Kieburg_Rigol_Vidmar_2022,PhysRevX.14.031014,fava2024}, tripartite mutual information~\cite{Hosur2016, Lensky_2019, Sunderhauf2019}, and deep thermalization~\cite{PRXQuantum.4.010311, Ho_Choi_2022, Ippoliti_2023, Bhore_Desaules_2023, Kaneko_Iyoda_2020, Ippoliti2022solvablemodelofdeep} characterize chaos via the onset of universal random behavior by comparing with random matrices and Haar ensembles, which provide a benchmark for randomness~\cite{Roberts2017-en}. Yet, there is no unique way of probing the randomness of a quantum system. As such, quantum chaos does not have an agreed-upon definition and its quantifiers are varied and inconsistent. Dynamical probes of chaos like the Loschmidt echo and out-of-time ordered correlator (OTOC) have also been studied extensively without relying on comparisons with randomness, but have been shown to fail in a variety of cases~\cite{PhysRevLett.118.086801, Xu2020, PhysRevB.107.235421, Balachandran2022,
dowling_scrambling_2023}.

In contrast, classical chaos is a well-understood, deterministic phenomenon defined by the extreme sensitivity of a dynamical system to small perturbations. Intuitively, this results in complex, unpredictable time evolution whenever the phase space is coarse-grained, as it is in practice for any realistic observer. This temporal complexity is captured by the (Kolmogorov-Sinai) dynamical entropy---the Shannon entropy of a multi-time probability distribution over coarse-grained phase space---which quantifies the loss of information about a dynamical system over time~\cite{q_dyn_sys_Alicki, Fabio_Benatti_2009, schuster2006deterministic}. The growth rate of classical dynamical entropy unambiguously distinguishes random, chaotic, and non-chaotic dynamics~\cite{schuster2006deterministic} and has even been used to study chaos in semi-classical descriptions of quantum systems~\cite{bianchi2018, hallam2019}.

Alicki and Fannes have defined a quantum version of the dynamical entropy, which reduces to the classical dynamical entropy in an appropriate limit~\cite{lindblad_dyn_ent,Slomczynski1994-ns, Alicki1994-dc, Fabio_Benatti_2009}. Instead of trajectories in a coarse-grained phase space, this quantum dynamical entropy (QDE) quantifies the irreducible complexity of a sequence of measurement outcomes, minimized over all measurement bases. This sequential measurement defines a quantum stochastic process that is described mathematically by a process tensor~\cite{processtensor,Milz2020kolmogorovextension,milz_quantum_2021}: the most general description of a quantum system under repeated interventions. The process tensor also underpins important recent results in the theory of open quantum systems, including state-of-the-art simulation methods~\cite{strathearn2018,jorgensen_exploiting_2019,PhysRevA.102.052206, Cygorek_2022,dowling_trees} and an operational definition of quantum non-Markovianity~\cite{processtensor}.

QDE has been studied for single-particle systems~\cite{Alicki_kicked_1996,q_dyn_sys_Alicki, Alicki_2004_deco} and, more recently, quantum many-body models~\cite{Cotler2018,dowling_operational_2023}, with a similar quantity being studied under the guise of decoherent histories~\cite{Strasberg2024first,wang2025decoherenceh}. Unlike its classical counterpart, however, QDE may not unambiguously distinguish chaotic from non-chaotic dynamics in a many-body setting. Scaling of dynamical entropy with the number of measurements has been shown to have a maximal growth rate, not only for highly chaotic systems such as the Sachdev-Ye-Kitaev model but also regular dynamics including free fermions in the thermodynamic limit~\cite{Cotler2018} and a class of models called Lindblad-Bernoulli shifts~\cite{Lindblad1979-tm, dowling_operational_2023}. This can be explained as resulting from a choice of interventions that act locally on a subsystem: any information lost to the rest of the system---a feature that is not unique to chaotic dynamics---will cause dynamical entropy to grow.

Motivated by these deficiencies of the QDE, we develop a framework to analyze chaos in quantum stochastic processes that captures both the unpredictability of the observed dynamics and the complexity of the conditional quantum states. Our approach is based on the pure process tensor~\cite{dowling_operational_2023}, which describes all temporal and spatial correlations of a sequentially measured quantum many-body system, as detailed in Sec.~\ref{Background}. Because bipartite entanglement in a pure state is equivalent to the entropy of the subsystems, the QDE is nothing but the entanglement entropy between the temporal and spatial parts of this pure process tensor, as discussed in Refs.~\cite{Alicki_kicked_1996,Cotler2018, dowling_operational_2023}.
In order to capture the complexity of the output state, a natural next step is to consider entanglement across other bipartitions of the process that mix its spatial and temporal aspects: this spatiotemporal entanglement (STE) was introduced in Ref.~\cite{dowling_operational_2023}. It was argued there, that by keeping some of the final (spatial) state together with the temporal parts while tracing the rest, STE should also be sensitive to how correlations spread throughout the final state, i.e.~the information scrambling of the dynamics. However, neither QDE nor STE have been studied in detail for short-ranged lattice models, where rich phenomenology can emerge from the competition between inertia, interactions, and disorder.

Our primary contributions in this work are twofold. First, we perform an extensive numerical study of the pure process tensor generated by local interventions on paradigmatic spin-chain models, spanning non-interacting, interacting-integrable, chaotic, and many-body localized regimes. In all cases, we use exact diagonalization to perform time evolution; see Sec.~\ref{sec:model} for details of the models considered. Our work is the first \textit{in silico} experiment to systematically investigate how a quantum stochastic process is impacted by the presence or absence of ergodicity in the underlying Hamiltonian dynamics. As an auxiliary technical contribution, we show how to exploit symmetries to reach large system sizes in exact diagonalization studies of quantum stochastic processes. In particular, we focus on models that conserve particle number and use a restricted basis of interventions to construct a process tensor within a single $\mathrm{U}(1)$ symmetry sector. Our construction follows similar principles to those presented in Ref.~\cite{Milz2018}, and may be of use in pushing the boundaries of other numerical techniques based on the process tensor framework~\cite{strathearn2018,jorgensen_exploiting_2019,PhysRevA.102.052206,Cygorek_2022,dowling_trees}. Very recently, Ref.~\cite{schultz2025} also considered symmetry restrictions on the QDE, albeit for single-body quantum dynamics.

Our results for QDE are consistent with previous work~\cite{Cotler2018,dowling_operational_2023}, implying maximal growth rate in the thermodynamic limit for all systems that are not localized  (Sec.~\ref{results:QDE}). We further link this behavior of QDE to a strong suppression of non-Markovian temporal correlations, which is especially evident in chaotic systems. At long times, the QDE equilibrates to a value depending on the ergodicity of the model, but this is a finite-size effect. Conversely, the growth of STE is suppressed for non-interacting or localized systems, but grows rapidly for both chaotic and interacting integrable dynamics, making the latter difficult to distinguish using either QDE or STE in the thermodynamic limit (Sec.~\ref{results:STE}).

To remedy this, we introduce our second major contribution: the projected process ensemble (PPE). This is an ensemble of pure output states of the quantum stochastic process, conditioned on a given sequence of interventions, as detailed in Sec.~\ref{sec:PPE}. We show that the QDE and STE can be recovered from the first statistical moment of this ensemble. Higher-order moments characterize how entanglement within the system depends on the intervention, thus capturing both the spatial and temporal complexity of the quantum stochastic process. Unlike QDE and STE, these higher moments depend on the choice of local intervention, and we consider both local unitary operations and local measurements. 

We numerically compute the resulting PPE entanglement distributions and compare them with the Haar ensemble~\cite{FigueroaRomero_Modi_Pollock_2019,PhysRevLett.71.1291, PhysRevD.100.105010, PhysRevLett.124.050602} in Sec.~\ref{results:HOM}. Our finite-size scaling analysis indicates that the higher PPE moments clearly distinguish chaotic from integrable dynamics, even at large system size.
In particular, we find that the mean entanglement of the PPE for chaotic systems is close to the Haar-random value, while the variance vanishes exponentially with system size, in stark contrast to the behavior of non-chaotic systems. In simple terms, our results imply that any sequence of local unitary interventions on a chaotic many-body system will generate the maximal entanglement allowed by symmetry and locality constraints.
However, this effect is reduced when considering non-unitary interventions (i.e., local measurements), mirroring the phenomenology of measurement-induced phase transitions~\cite{Skinner2019, PhysRevB.100.134306, PhysRevB.99.224307,PhysRevB.98.205136} where measurement backaction inhibits the proliferation of entanglement within the many-body system.

Our approach can be seen as a temporal version of the projected state ensembles recently introduced to study deep thermalization~\cite{choi_preparing_2023, PRXQuantum.4.010311,Ho_Choi_2022,Ippoliti_2023}. In fact, objects similar to the process tensor have been used to define projected ensembles in the past. In Ref.~\cite{Ippoliti2022solvablemodelofdeep}, a state ensemble is constructed from the conditional states in time after projecting onto the output states of a process. Conversely, here we construct the PPE from the output states of a process, conditioned on a particular multi-time measurement or control operation. 

The entanglement properties studied here should also not be confused with temporal entanglement~\cite{sim_complex_Filippov_19, PhysRevResearch.6.033021, boucomas2024}, i.e.~the bipartite entanglement within a vectorized marginal process tensor describing temporal degrees of freedom only. The latter has been shown to have distinctive scaling behavior depending on the chaoticity of the model~\cite{Lerose2021, Giudice2022, PhysRevX.13.041008}, with important implications for the efficiency of numerical algorithms based on temporal matrix-product states~\cite{Banuls2009,  jorgensen_exploiting_2019,PhysRevA.102.052206, Lerose2021b, Cygorek_2022, Carignano2024}. Interestingly, here we find that temporal correlations in the full process tensor become vanishingly small in chaotic systems, at least for sufficiently long time intervals between interventions. This contrasting behavior in comparison to the volume-law scaling of temporal entanglement in chaotic circuits~\cite{PhysRevX.13.041008} deserves further scrutiny, and could even inform the development of new simulation methods to capture relevant temporal correlations efficiently by an appropriate coarse-graining~\cite{White2018,Rakovsky2022, dowling_trees}. We discuss this and other possible avenues for future work in Sec.~\ref{sec:discussion}.

\section{Framework}
\label{Background}

\subsection{Pure Process Tensor}
\label{sec:pure_pt}

A classical stochastic process is the joint multi-time probability distribution for a random variable. Its quantum generalization is a process tensor~\cite{OperationalQDynamics, milz_quantum_2021}: a multi-time density matrix. Process tensors enable the formulation of rigorous definitions of Markovianity \cite{processtensor, processtensor2, milz_completely_2019, taranto2019memory, Taranto2019FiniteMarkov, Markovorder1, White2020, White2021, PRXQuantum.3.020344, Zambon_2024} and have been used to develop efficient algorithms for simulating the dynamics of quantum systems \cite{jorgensen_exploiting_2019, PhysRevA.102.052206, 
 PRXQuantum.4.020310, Cygorek_2022, dowling_trees}. Previous works have also used process tensors to understand quantum chaos \cite{zonnios_signatures_2022, dowling_operational_2023} and equilibration \cite{Dowling2023relaxationof,finitetime, FigueroaRomero_Modi_Pollock_2019, FigueroaRomero_Pollock_Modi_2021}. In this section, we follow Ref.~\cite{dowling_operational_2023} to define the pure state process tensor which is a pure, multi-time quantum state. Entanglement properties of this state define the quantum dynamical entropy (QDE)~\cite{Alicki1994-dc} and spatiotemporal entanglement (STE)~\cite{dowling_operational_2023}. We then introduce the projected process ensemble which promotes these quantities to random variables. In other words, we show exactly how these seemingly unrelated ideas stem from the same fundamental description of the spatiotemporal process.

Consider an initial pure state of the system and environment $|\psi_{R}\rangle \in \mathcal{H}_{SE} \equiv \mathcal{H}_R$ and dynamics described by a unitary operator $U$. We have denoted the total system and environment Hilbert space as $\mathcal{H}_R$ which will be referred to as the remainder space. This space has dimension $d_R = d_Sd_E$, where $d_S$ and $d_E$ are the Hilbert space dimension of system and environment, respectively. The interventions are given by Kraus operators $A_{x}$ which are an instrument decomposition of a completely positive, trace-preserving (CPTP) map such that $\sum^r_{x = 1} A_x^{\dagger} A_x = \id_{R}$. We assume that these Kraus operators are local on the system Hilbert space, such that $A_{x} = A^{(S)}_{x}\otimes \id^{(E)}$. The operator $A_x$ allows us to compute both the probability of measuring outcome $x$ as well as the output state upon obtaining that measurement outcome. We impose an additional condition that the operators are orthogonal with respect to the Hilbert-Schmidt inner product, $\tr(A^{\dagger}_{x} A_{y}) = \delta_{xy}$, and span the full space of local operators on $S$ such that they are informationally complete.

Using these operators, the (unnormalized) output state of a quantum stochastic process from the action of a sequence of interventions $A_{x_{1}}, ..., A_{x_{n_B}}$ at times $\vec{t} = \{t_1, t_2 ..., t_{n_B}\}$ is given by
\begin{equation}
\label{eq:output_state}
    |\ups_{R|\vec{x}}\rangle = A_{x_{n_B}}(t_{n_B})... A_{x_{1}}(t_{1})|\psi_{R}\rangle,
\end{equation}
with  $A_{x_{i}}(t_{i}) = U^{\dagger}(t_{i}) A_{x_{i}} U(t_{i})$ and $\vec{x} = \{x_1, ..., x_{n_B}\}$. The action of these interventions have different interpretations depending on whether they are deterministic or non-deterministic. Deterministic interventions are proportional to unitary operators, i.e.~$A_{x_i}= U_{x_i}/\sqrt{d_S}$ where $U_{x_i}$ is unitary, such that all output states are subnormalized as $\langle \ups_{R|\vec{x}}|\ups_{R|\vec{x}}\rangle = 1/d_{S}^{n_B}$. Non-deterministic interventions, given by projection operators, describe the effect of obtaining a sequence of measurement outcomes $\vec{x}$ at times $\vec{t}$ with the probability expressed simply as the norm of the final, subnormalized state $p_{\vec{x}} = \langle\ups_{R|\vec{x}}|\ups_{R|\vec{x}}\rangle$. In later sections, we will compare the effect of deterministic and non-deterministic operators on the process and the quantities we use to probe chaos.

To construct the process tensor we represent the measurement operators using the Choi-Jamio\l kowski isomorphism \cite{wilde2020, Watrous_2018, nielsen_chuang_2010}. Using linearity and complete positivity, the Choi-Jamio\l kowski representation of the interventions is constructed from their action on one part of a maximally entangled state:
\begin{equation}
    |x_j\rangle = A^{(S)}_{x_j} \otimes \mathds{1}_{S} \sum_{k}|k\rangle_{S}\otimes |k\rangle_{S} = \sum_{k} A^{(S)}_{x_i}|k\rangle_{S} \otimes |k\rangle_{S}.
\end{equation}
Thus, $A_{x_j}$ is mapped to a pure state in a doubled Hilbert space, $|x_j\rangle \in \mathcal{H}^{i o}_S$, which corresponds to the input ($i$) and output ($o$) spaces of the operator. Using this mapping, we can construct pure states corresponding to different multi-time interventions,
\begin{equation}
\label{Eq:butterfly_space}
    |\vec{x}\rangle = |x_{1}\rangle \otimes ...\otimes |x_{n_B}\rangle \in \mathcal{H}_{B}.
\end{equation}
Here, we call the full multi-time intervention space, $\mathcal{H}_{B}=\mathcal{H}^{i_{1}o_1}_S(t_1)\otimes ... \otimes \mathcal{H}^{i_{n_B} o_{n_B}}_S(t_{n_B})$, the butterfly space, for reasons which will become apparent when we describe different realizations of the (multi-time) butterfly effect in the next section.

We can construct a multi-time pure state using a generalized Choi-Jamio\l kowski isomorphism applied to the process tensor---the CPTP superchannel which maps local interventions made on the system at multiple times to the corresponding output state. The pure process tensor is given by the following,  
\begin{equation}
\label{eq:pure_process_tensor}
|\Upsilon\rangle = \sum_{\vec{x}} \underbracket[0.8pt]{|\ups_{R|\vec{x}}\rangle}_{\text{Spatial (R)}} \! \otimes \! \underbracket[0.8pt]{|\vec{x}\rangle}_{\text{Temporal (B)}}.
\end{equation}
We have labelled the spatial and temporal components of the pure process tensor, which are elements of the Hilbert spaces  $\mathcal{H}_R$ and $\mathcal{H}_B$, respectively. The process tensor in Eq.~\eqref{eq:pure_process_tensor} has the following properties,
\begin{equation}
    \langle \Upsilon|\Upsilon\rangle = 1, \; \; \; \text{and} \; \; \; \langle\vec{x}|\vec{y}\rangle = \delta_{\vec{x}\vec{y}}.
\end{equation}
These ensure that the pure process tensor $|\Upsilon\rangle$ is a genuine quantum state and that distinct interventions $\vec{x}$ correspond to orthonormal basis states in $\mathcal{H}_B$.

Physically, the butterfly space $\mathcal{H}_B$ can be interpreted as a quantum register that coherently records the possible outputs of the process. Each butterfly state $|\vec{x}\rangle$, pertaining to a distinct sequential intervention $\vec{x}$ on the local subsystem, is paired with its corresponding output state, i.e.~the conditional state $|\Upsilon_{R|\vec{x}}\rangle$ of the global system. The pure process tensor therefore encodes both quantum and classical temporal correlations, while also allowing one to obtain the output state for any sequence of interventions on the local subspace. Crucially, the interventions are informationally complete, meaning that the butterfly states $|\vec{x}\rangle$ form a complete basis. Therefore, the output states and probabilities for any other sequential intervention (e.g., measurements in a different basis) can also be obtained by an appropriate unitary rotation in the butterfly space.

Following Ref.~\cite{dowling_operational_2023}, our construction of the process tensor assumes pure initial states and unitary dynamics, which are most natural to consider when studying deterministic quantum chaos. However, more general constructions are also possible, such as for non-unitary dynamics and/or mixed initial states, leading to a process tensor whose Choi-Jamio\l kowski representation is a mixed quantum state~\cite{OperationalQDynamics, milz_quantum_2021}. This conventional (mixed) process tensor is recovered in our approach by tracing out the remainder space. The result is a density matrix on the butterfly space [Eq.~\eqref{eq:reduced_pt} below] whose diagonal elements encode the probabilities of obtaining specific outcome sequences, while the off-diagonal elements capture the degree of orthogonality between output states for distinct sequences. As we will see, these off-diagonal terms provide a natural probe of a quantum analogue of the butterfly effect.

\begin{figure*}[t]
    \centering
    \includegraphics[width=1\textwidth]{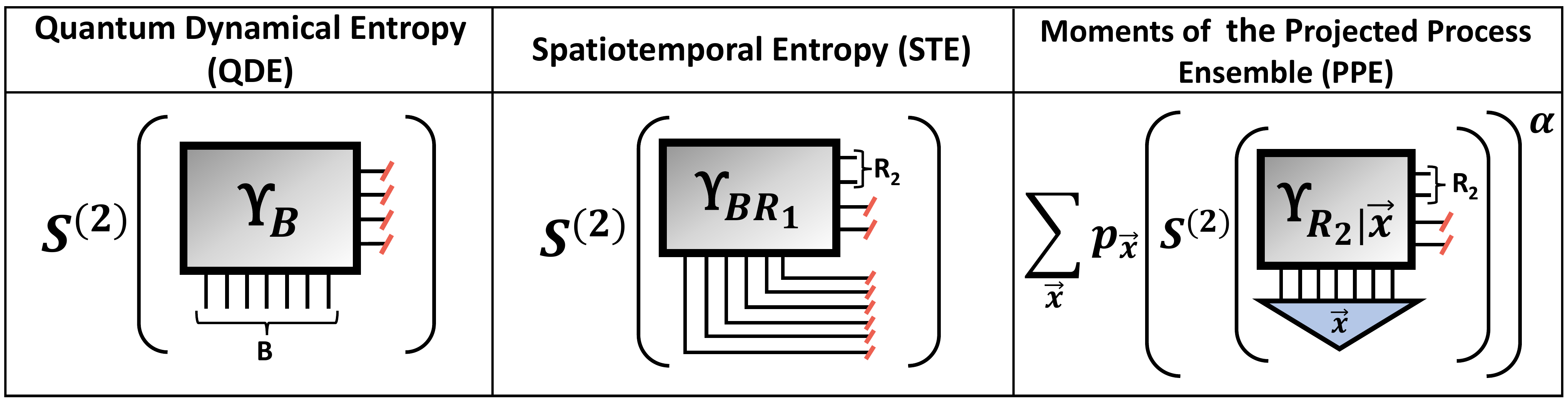}  
    \caption{Quantities considered in this work using the R\'enyi-2 entropy, denoted as $S^{(2)}$, to quantify chaotic dynamics using the entanglement properties of the process tensor. Quantum dynamical entropy (QDE), defined in Sec.~\ref{sec:DE}, quantifies the entanglement between the past interventions and the output state of the process. In Sec.~\ref{sec:STE}, the STE is defined and extends upon the QDE by including information about the complexity of the output states. Finally, we define the mean and the variance---corresponding to $\alpha = 1$ and $\alpha = 2$, respectively---of the projected process ensemble in Sec.~\ref{sec:PPE}. These quantities provide fine-grained probes of the output state complexity.}
    \label{Fig:idea}
\end{figure*}

\subsection{Figures of Merit}
\label{sec:figs_of_merit}
The pure process tensor allows us to represent dynamical properties of quantum systems using information-theoretic methods for pure states. We now discuss a series of entropic quantities which will capture signatures of quantum chaos with varying degrees of sensitivity.

\subsubsection{Quantum Dynamical Entropy}
\label{sec:DE}

The first quantity we consider is the QDE. Classically, dynamical entropy is an entropy rate from repeated measurements of a dynamical systems. In quantum systems, there is no unique definition of the dynamical entropy; however, a well-known version is given by the Alicki-Fannes dynamical entropy~\cite{Pechukas1982, Alicki1994-dc,Slomczynski1994-ns, q_dyn_sys_Alicki, Fabio_Benatti_2009}. This quantity has been studied analytically in a class of simple models called Lindblad-Bernoulli shifts~\cite{q_dyn_sys_Alicki, Alicki_2004_deco} and has been numerically analyzed in free-fermion, Sachdev-Ye-Kitaev~\cite{Cotler2018}, and kicked rotor models~\cite{Alicki_kicked_1996, Alicki_2004_deco}, where it has been shown to display some signatures of chaos.

The QDE can be expressed as the entanglement entropy across the $B:R$ bipartition of the pure process tensor~\cite{dowling_operational_2023}. Using Eq.~\eqref{eq:pure_process_tensor}, the reduced state on $\mathcal{H}_B$ is given by
\begin{equation}
\label{eq:reduced_pt}
    \ups_{B} = \sum_{\vec{x}, \vec{y}} \langle \ups_{R|\vec{y}} |\ups_{R|\vec{x}}\rangle |\vec{x}\rangle \langle \vec{y}|.
\end{equation}
We then use the R\'enyi-2 entropy to quantify the entanglement between $B$ and $R$:
\begin{equation}
\label{eq:dyn_ent}
    E(B:R) = S^{(2)}(\ups_{B}) = -\log\left[\tr(\Upsilon_{B}^2)\right].
\end{equation}
We will refer to Eq.~\eqref{eq:dyn_ent} as the QDE throughout. Due to the purity of the process tensor, the R\'enyi-2 entropy of the reduced state $\ups_{R}$ would give identical results. This entropy is directly interpreted as a measure of entanglement between the spatial and temporal components of the pure process tensor. In Appendix~\ref{Appendix:ALF_vs_PT}, we discuss the differences and similarities between the QDE in Eq.~\eqref{eq:dyn_ent} and the Alicki-Fannes entropy.

QDE has been interpreted as a probe of the butterfly flutter fidelity:~a strong, non-local sensitivity of quantum processes to small, local perturbations made at one or more times in the past. Here, sensitivity means orthogonality of the output states of a process for different interventions. When the QDE in Eq.~\eqref{eq:dyn_ent} is (nearly) maximal, all orthogonal interventions, $|\vec{x}\rangle$ and $|\vec{y}\rangle$ such that $\langle \vec{x}|\vec{y}\rangle = 0$, with output states, $|\ups_{R|\vec{x}}\rangle$ and $|\ups_{R|\vec{y}}\rangle$, will be (nearly) orthogonal~\cite{dowling_operational_2023},
\begin{equation}
\langle\ups_{R|\vec{x}}|\ups_{R|\vec{y}}\rangle \approx 0.
\end{equation}
The orthogonalization of the outputs for different interventions is analogous to the classical butterfly effect where the dynamics is sensitive to small perturbations. 

Given this interpretation, we expect that chaotic dynamics will have maximal or near-maximal dynamical entropy and non-chaotic dynamics such as integrable or many-body localized models will produce submaximal dynamical entropy. In Sec.~\ref{sec:results}, we will confirm this intuition through a detailed numerical analysis of finite-sized systems. However, maximal growth of dynamical entropy can also be found for certain integrable models, as we now discuss.

\subsubsection{Spatiotemporal Entanglement}
\label{sec:STE}

A class of non-chaotic models called Lindblad-Bernoulli shifts have been shown to display maximal growth rate of the QDE, despite the regularity of their dynamics~\cite{dowling_operational_2023}. Free fermions have also been shown to exhibit similar growth of the QDE in the thermodynamic limit and after sufficiently long timescales~\cite{Cotler2018}. Given the inability of the QDE to fully characterize chaos, we now define the spatiotemporal entanglement (STE), introduced in Ref.~\cite{dowling_operational_2023}, which can distinguish Lindblad-Bernoulli-type dynamics from chaotic dynamics.

The process-tensor formulation of the QDE allows for a natural generalization to the spatiotemporal setting. We are free to choose arbitrary bipartitions of the form $B_1 R_1:B_2 R_{2}$, allowing us to define the \emph{spatiotemporal entanglement} entropy (STE). Denoting the reduced state on $B_2 R_{2}$ as $\ups_{B_2 R_{2}} =\tr_{B_1 R_{1}}\left[|\ups\rangle\langle \ups|\right]$, and using the R\'enyi-2 entropy to quantify entanglement, we define the STE as
\begin{equation}
\label{eq:st_entropy}
    E(B_1R_1:B_2 R_2) = S^{(2)}(\ups_{B_2 R_{2}}) = - \log(\tr(\ups_{B_{2} R_{2}}^2)).
\end{equation}
To optimally distinguish chaotic and non-chaotic dynamics using this quantity, one should minimize over the choice of bipartition~\cite{dowling_operational_2023}; however, due to the difficulty in doing this, we instead restrict ourselves to bipartition of the form $BR_1:R_2$.

The STE includes information about the complexity of the output states of the process. We can therefore interpret this quantity as a more refined QDE which probes the information scrambling ability of the dynamics---analogous to other known methods of probing chaos such as out-of-time ordered correlators~\cite{Shenker_Stanford_2014, Maldacena_Shenker_Stanford_2016, Swingle2016, Roberts2016, swingle_unscrambling_2018, xu_scrambling_2024}, tripartite mutual information~\cite{Hosur2016, Lensky_2019, Sunderhauf2019}, and subsystem information capacity~\cite{chen2024subsystem}. Given the ability of this quantity to accurately characterize the non-chaoticity of Lindblad-Bernoulli shifts, we expect it to be more capable of distinguishing chaotic and non-chaotic dynamics. Again, this entropy is also interpreted as an entanglement entropy between subspace $BR_1$ and $R_2$. However, in general, the STE requires minimizing over a bipartition which is generically difficult to perform. Without performing this minimization, we are not be able to rule out the existence of a bipartition which may display signatures of chaos for non-chaotic dynamics.

\subsubsection{Projected Process Ensemble}
\label{sec:PPE}

The QDE quantifies the sensitivity of the output states for different interventions by their orthogonality. As discussed, this does not fully capture chaotic behavior and can be spoofed by non-chaotic dynamics. The STE characterizes both sensitivity to interventions and includes information about the complexity of the output states. This is sufficient to distinguish some counterexamples which spoof the QDE, but may not be unambiguous. In this section, inspired by deep thermalization~\cite{PRXQuantum.4.010311}, we define the projected process ensemble which will enable us to define higher order probes of complexity of the output states, and allow us to quantify chaotic dynamics on a more fine-grained level. 

To extend deep thermalization to the spatiotemporal domain, we consider the pure process in Eq.~\eqref{eq:pure_process_tensor}. For a fixed, orthogonal basis of interventions, $|\vec{x}\rangle \in \mathbf{X}$, we can obtain the conditional output states of the process by projecting onto the butterfly space $\langle \vec{x}|\ups\rangle = |\ups_{R|\vec{x}}\rangle$. We can then define an ensemble comprised of these output states, $|\ups_{R|\vec{x}}\rangle$, and the probabilities of obtaining the sequence of measurement outcomes, $p_{\vec{x}} = \langle \ups_{R|\vec{x}}|\ups_{R|\vec{x}}\rangle$. This ensemble, which we refer to as the \textit{projected process ensemble} (PPE), is given by,
\begin{equation}
\label{eq:PPE}
    \mathcal{E} \equiv \{p_{\vec{x}}, |\tilde{\ups}_{R|\vec{x}}\rangle\}_{|\vec{x}\rangle \in \mathbf{X}},
\end{equation}
where we normalized the output states such that $|\ups_{R|\vec{x}}\rangle \rightarrow |\tilde{\ups}_{R|\vec{x}}\rangle = \frac{1}{\sqrt{p_{\vec{x}}}}|\ups_{R|\vec{x}}\rangle$.

---

In general, projected ensembles have been used extensively in the study of quantum chaos. In this context, quantum chaos is described by deep thermalization~\cite{PRXQuantum.4.010311, choi_preparing_2023}---whereby ensembles generated by chaotic evolution combined with projective measurements over a subspace of the Hilbert space appear to be maximally random as defined by the Haar ensemble. Deep thermalization has been studied analytically and numerically for a variety of models \cite{Kaneko_Iyoda_2020, Ho_Choi_2022, Ippoliti_2023, Bhore_Desaules_2023, Chan_2024}, but has primarily focused on single-time, spatial projected ensembles. Alternative methods for temporal projected ensembles have been considered; for instance, in Ref.~\cite{Ippoliti2022solvablemodelofdeep}, the design time of an exactly solvable model is studied using a process tensor-like object whereby the ensemble comprises states in the butterfly space conditioned on projections made on the outputs of the process. For a fixed, orthogonal basis of projections on the remainder space, $|\vec{z}\rangle \in \mathbf{Z}$, the conditional states are defined as $|\ups_{B|\vec{z}}\rangle = \sum_{\vec{x}} \langle \vec{z}|\ups_{R|\vec{x}}\rangle \otimes |\vec{x}\rangle$. These states define an ensemble, $\mathcal{E}_B \equiv \left\{p_{\vec{z}}, \frac{1}{\sqrt{p_{\vec{z}}}}|\ups_{B|\vec{z}}\rangle \right\}$, with $p_{\vec{z}} = \text{tr}_{R}\left[\ups_{R} |z\rangle \langle z|_R\right]$. Ensembles have also been constructed by randomizing the duration of time evolution~\cite{PhysRevX.14.041051} and using the eigenstates of chaotic many-body Hamiltonians~\cite{PhysRevX.14.031014}. These approaches contrast the ensemble considered in this work which is constructed by projecting onto the butterfly space with a fixed basis of interventions, as shown in Eq.~\eqref{eq:PPE}. This can be understood as looking at the set of output states of the process conditioned on obtaining a particular sequence of measurement outcomes.

From this ensemble, higher-order moments can be constructed; for instance, the $k$-th order moment of the PPE is given by
\begin{equation}
\label{eq:k_moment}
    \ups_{R}^{(k)} = \sum_{\vec{x}} p_{\vec{x}} |\tilde{\ups}_{R|\vec{x}}\rangle \langle \tilde{\ups}_{R|\vec{x}}|^{\otimes k}.
\end{equation}
The QDE and the STE can be expressed in terms of the first moment as $\nolinebreak{E(B:R) = S^{(2)}(\ups^{(1)}_{R})}$ and $\nolinebreak{E(BR_1:R_2) = S^{(2)}\left(\tr_{R_1}[\ups^{(1)}_{R}]\right)}$. By studying higher-order moments, we expect to be able to more accurately distinguish chaotic from non-chaotic dynamics.

The first probe we consider is the average bipartite entanglement of the output states of the process. For output state  $|\tilde{\ups}_{R|\vec{x}}\rangle$, the bipartite entanglement is given by
\begin{equation}
    S_{\vec{x}} = -\log\left[\tr[\ups_{R_2|\vec{x}}^2]\right],
\end{equation}
with $\ups_{R_2|\vec{x}} = \text{tr}_{R_1}\left[|\tilde{\ups}_{R|\vec{x}}\rangle\langle \tilde{\ups}_{R|\vec{x}}|\right]$. Averaging this quantity over the PPE yields
\begin{equation}
    \langle S_{\vec{x}}\rangle_{\mathcal{E}} = \sum_{\vec{x}}p_{\vec{x}}S_{\vec{x}}.
\end{equation}
The R\'enyi-2 entropy requires access to two copies of the output states, so it is a probe of the second moment of the PPE ensemble. The second probe we consider is the standard deviation of the entanglement entropy,
\begin{equation}
    \Delta_{\mathcal{E}} S_{\vec{x}} = \sqrt{\langle S_{\vec{x}}^2\rangle_{\mathcal{E}} - \langle S_{\vec{x}}\rangle_{\mathcal{E}}^2},
\end{equation}
which is a probe of the fourth moment of the PPE ensemble. We will show through numerical analysis that these quantities can distinguish chaotic from non-chaotic dynamics and that distinct system-size scaling behavior occurs which is not observed in the QDE or STE, making the moments of the PPE a more robust approach for characterising chaos.

\subsection{Summary}

We now summarise the quantities defined above and provide historical context and physical intuition behind them. A schematic illustration of all these quantities is shown in Fig.~\ref{Fig:idea}.

The QDE is the first probe considered in this work. This quantity was initially defined by Lindblad in Ref.~\cite{Lindblad1979-tm}, and was then rediscovered by Alicki and Fannes in Ref.~\cite{Alicki1994-dc} (see Sec.~\ref{Appendix:ALF_vs_PT} for details of this definition). Later, the QDE was shown to be equivalent to the bipartite entanglement between the spatial and temporal regions of the process tensor~\cite{Alicki_kicked_1996}. Physically, the correlations between the measurement record $\vec{x}$ (temporal region) and the output states $|\Upsilon_{R|\vec{x}}\rangle$ (spatial region) can be interpreted as the sensitivity of the outputs to different interventions made in the past—a quantum analogue of the butterfly effect. Therefore, a large QDE signals strong sensitivity of the outputs of the process to different interventions, which is consistent with chaotic dynamics. However, this behaviour is not unique to chaos, as shown in Refs.~\cite{Cotler2018, dowling_operational_2023}. In Sec.~\ref{results:QDE}, we similarly conclude that the scaling of the QDE alone is insufficient to distinguish chaotic from non-chaotic dynamics.

The STE is the second probe considered in this work. Introduced in Ref.~\cite{dowling_operational_2023}, it generalises the QDE by quantifying correlations across more general bipartitions of the pure process tensor that mix space and time. These correlations probe both the sensitivity of the output states to past interventions and the spatial complexity of those outputs. It has been shown that such correlations can distinguish certain non-chaotic dynamics that mimic chaotic QDE growth~\cite{dowling_operational_2023}. We show in Sec.~\ref{results:STE} that this provides a sharper diagnostic of chaos, yet still struggles to distinguish chaotic and non-chaotic dynamics at large system size.

Finally, we introduce the PPE as a novel tool to diagnose chaos. This is an ensemble of output states conditioned on a fixed basis of local interventions at multiple times. The statistical moments of the PPE encompass the QDE and the STE, while higher-order moments can be used to probe the complexity and randomness of the dynamics at a more fine-grained level. Agreement between these moments and those of the Haar ensemble implies a temporal form of deep thermalisation, which we observe numerically in Sec.~\ref{results:HOM}. The PPE is constructed from forward-in-time evolution only, distinguishing it from many other chaos measures that probe both forward and backward time evolution.

In the following sections, we analytically and numerically study these quantities and use them to understand chaotic behaviour.



\begin{table*}[ht]
    \centering
    \begin{tabular}{|c||c|c|c|c|}
    \hline
     \textbf{Model} & \textbf{Hamiltonian} & \textbf{Classification} & \textbf{Deterministic} & \textbf{Non-Deterministic}\\
    \hhline{|=#=|=|=|=|}
    \textbf{XXZ} & $J \sum_{i=1}^L\left(\sigma_i^x \sigma_{i+1}^x+\sigma_i^y \sigma_{i+1}^y+\Delta \sigma_i^z \sigma_{i+1}^z\right)$ & Interacting Integrable & $\bigl\{\frac{\id_1}{\sqrt{2}}, \frac{\sigma^z_1}{\sqrt{2}}\bigr\}$ & $\bigl\{|0\rangle \langle 0|, |1\rangle \langle 1|\bigr\}$ \\
    \hline
    \textbf{XXZ with NNN} & $H_{\text{XXZ}}(J, \Delta) + \sum^L_{i = 1} h \sigma^z_{i} \sigma^z_{i + 2} + g\sigma^z_{1}$ & Chaotic & $\bigl\{\frac{\id_1}{\sqrt{2}}, \frac{\sigma^z_1}{\sqrt{2}}\bigr\}$& $\bigl\{|0\rangle \langle 0|, |1\rangle \langle 1|\bigr\}$\\
    \hline
    \textbf{IAA} & $-H_{\text{XXZ}}(J, \Delta) + 2\lambda \sum_{i=1}^L\cos(2\pi qi) \sigma^z_{i}$ & Chaotic/Localized & $\bigl\{\frac{\id_1}{\sqrt{2}}, \frac{\sigma^z_1}{\sqrt{2}}\bigr\}$ &  $\bigl\{|0\rangle \langle 0|, |1\rangle \langle 1|\bigr\}$ \\
    \hline
     \textbf{Free Fermion} & $\sum^L_{i,j=1} J_{ij} c^{\dagger}_{i} c_j$& Integrable & $\bigl\{\frac{\id_1}{\sqrt{2}},  \frac{c^{\dagger}_{1} c_{1} - c_{1} c^{\dagger}_1}{\sqrt{2}}\bigr\}$ & $\bigl\{c^{\dagger}_{1} c_{1}, c_{1} c^{\dagger}_1 \bigr\}$\\
    \hline
    \end{tabular}
    \caption{One-dimensional spin-chain and fermionic, many-body Hamiltonians considered in this work with corresponding choice of intervention basis. These models are characterized as chaotic or non-chaotic based on whether or not they exhibit ergodic behavior. In all cases, we consider periodic boundary conditions. For the interacting integrable XXZ model, which is denoted as $H_{\text{XXZ}}(J, \Delta)$, we take parameters $J = 1$, $\Delta = 0.55$, and for the chaotic case we take $h = 0.6$, and $g = 0.1$. The interacting Aubry-Andr\'{e} (IAA) model takes $J = 1$, $\Delta = -1$, $q = \frac{2}{\sqrt{5} + 1}$, with $\lambda = 1$ or $\lambda = 5$ for the chaotic and localized cases, respectively. The free fermion model has $J_{ij} = J(\delta_{j, i+1} + \delta_{i+1, j})$ with $J = 1$. In each case, we consider a basis of interventions which preserves the particle number symmetry. When studying the PPE, we will compare deterministic and non-deterministic interventions which are shown in the final two columns.}
    \label{tab:model}
\end{table*}

\section{Models and Methods}
\label{sec:model}

In Sec.~\ref{sec:results} we numerically compute the QDE, the STE, and the mean and variance of the PPE for dynamics generated by chaotic and non-chaotic many-body Hamiltonians. Before discussing our numerical results, we first explain the properties of the different models considered and then discuss how we work with the symmetries in these models to improve the efficiency of our numerics.

\subsection{Many-Body Models}

The four models we study are detailed in Table~\ref{tab:model}. We describe below why we have selected these models and how they collectively offer insight for understanding quantum chaos.

The first model is the XXZ model, detailed in the first row of Table~\ref{tab:model}, which is an interacting integrable (Bethe ansatz integrable) model~\cite{Sutherland_models}. Integrability is an example of ergodicity breaking whereby the models have an extensive number of locally conserved quantities, so these models are considered to be non-chaotic. Interacting integrable models have been shown to display chaotic behavior in the presence of small perturbations~\cite{brenes_low-frequency_2020,PhysRevX.10.041017,santos_speck_2020}. Adding next-to-nearest neighbour interactions breaks most of the symmetries of the XXZ model, giving us a chaotic model, which is described in the second row of Table~\ref{tab:model}.

We also consider the interacting Aubry-Andr\'e (IAA) model shown in the third row of Table~\ref{tab:model}. This model is a deterministic, quasiperiodic model which is known to exhibit a chaotic and many-body localized regime~\cite{PhysRevB.87.134202, PhysRevLett.119.075702, PhysRevLett.121.206601, PhysRevB.100.104203, Swingle_2019}. Many-body localization is another form of ergodicity breaking, distinct from integrability, in which excessive disorder causes local integrals of motion to emerge that constrain the dynamics~\cite{PhysRevLett.111.127201, PhysRevB.90.174202, PhysRevLett.117.120402, PhysRevB.91.085425, RevModPhys.91.021001}. This type of ergodicity breaking is known to be more robust against Hamiltonian perturbations compared to interacting integrability. In this model, disorder is introduced by an on-site potential $2\lambda \cos(2\pi q i)$ with an irrational number $q$, which takes a different value on each site $i$ despite being generated by a deterministic equation. For small values of $\lambda$, this disorder will result in chaotic behavior. For sufficiently large values of $\lambda$, many-body localization will occur.

Finally, in the fourth row of Table~\ref{tab:model}, we consider a free fermion model: a non-interacting integrable model which is robustly non-ergodic, in the sense that its integrability is preserved by an arbitrary quadratic Hamiltonian perturbation. Similar to many-body localization and unlike interacting integrability, these models are less sensitive to perturbations and tend to exhibit more distinct non-chaotic behavior~\cite{PhysRevX.10.041017}. 

Throughout this work,  we simulate dynamics through exact diagonalization of the Hamiltonians using QuSpin~\cite{10.21468/SciPostPhys.2.1.003, 10.21468/SciPostPhys.7.2.020}. For the XXZ model, we take $J = 1$, $\Delta = 0.55$, $h = 0.6$, and $g = 0.1$. For the IAA model, we take $J = 1$, $\Delta = -1$. We set $q = \frac{2}{\sqrt{5} + 1}$ which is irrational, resulting in a quasiperiodic potential. The strength of the disorder is set by $\lambda$. We consider models with $\lambda = 1$ and $\lambda = 5$, corresponding to chaotic and localized models, respectively. For the free fermion model, we take $J_{ij} = J(\delta_{j, i+1} + \delta_{i+1, j})$ with $J = 1$. Time units such that $\hbar=1$ are used throughout.

\subsection{Symmetry-Restricted Process Tensor Construction}
\label{sec:symmetry_PT}

An important feature of all the Hamiltonians shown in Table~\ref{tab:model} is that they enjoy a $U(1)$ symmetry, leading to conservation of the excitation or particle number $ N = \sum_{i=1}^L \tfrac{1}{2}(\id_i+\sigma^z_i)$ (spins) or $N=\sum_{i=1}^L c_i^\dagger c_i$ (fermions). This symmetry results in degeneracies in the energy spectrum making it more difficult to analyze chaotic behavior. To study chaos in these models, we resolve this symmetry by considering a particle-number subspace with half filling. For all cases, we consider an initial N\'eel state,
\begin{equation}
\label{eq:neel}
    |\psi_{0}\rangle = |01\rangle^{\otimes \frac{L}{2}},
\end{equation}
which has $N=L/2$ excitations. Under time evolution with respect to the Hamiltonians in Table~\ref{tab:model}, this state will remain in Hilbert space $\mathcal{H}_{R}(N = L/2)$ such that the dimension of the remainder space is given by $d_{R} = {L\choose L/2}$.

The local interventions do not, in general, conserve particle number. Consider a fixed basis of local interventions $A_x \in \mathcal{M}$ obeying the properties discussed in Sec.~\ref{sec:pure_pt}, and having the following commutation relation with the particle number operator $N$, 
\begin{equation}
    \label{N_eigenops}
    [N, A_x] =  
\Delta n_{x} A_{x},
\end{equation}
where $\Delta n_x = 0,\pm 1$. These operators will change the particle number by $\Delta n_x$, i.e.~they either add one, subtract one, or leave the particle number unchanged. We can characterize the multi-time instruments by how they change the particle number; for instance, a set $\mathcal{M}_{1:n_B}(\Delta N)$ contains multi-time interventions, $|\vec{x}\rangle$, which change the overall particle number by 
\begin{equation}
    \Delta N = \sum^{n_B}_{i = 1} \tr\left(A_{x_i}^{\dagger}[N,A_{x_i}]\right) = \sum_{i=1}^{n_B} \Delta n_{x_i}.
\end{equation}
Using this basis and assuming we choose an initial state within a particle subspace, we can decompose the process tensor into different particle number subsectors,
\begin{equation}
\label{pure_process_decomposition}
    |\ups\rangle = \sum^{n_B}_{\Delta N = -n_B} \sum_{\vec{x} \in \mathcal{M}(\Delta N)} |\ups_{R|\vec{x}}\rangle \otimes |\vec{x}\rangle.
\end{equation}
Process tensors of this form occupy a multi-time Hilbert space $\mathcal{H}_{RB} = \bigoplus^{L}_{N = 0}\mathcal{H}_R(N) \otimes \mathcal{H}_B$ with $\dim[\mathcal{H}_{RB}] = d_{S}^L \times d_{S}^{2 n_B}$.

To more efficiently compute the process tensor, we will ensure that we remain within one particle-number sector by considering only a basis of local interventions $A_{x}$ which commute with the particle number operator $[A_x, N] = 0$. For instance, a complete basis of interventions at site $1$ in a spin lattice model is given by the set of Pauli matrices; however, $\sigma^x$ and $\sigma^y$ do not preserve the particle number symmetry since $[\sigma^{x,y}_1, N] \neq 0$. As a result, we take a reduced set of Pauli matrices,
\begin{equation}
    \{\mathds{1}_1, \sigma^x_1, \sigma^y_1, \sigma^z_1\} \rightarrow \{\mathds{1}_1, \sigma^z_1\}.
\end{equation}
Here, the symmetry sectors correspond to a group module decomposition of the Hamiltonian $H$, and the unitary maps, $\sigma_z$ and $\mathds{1}$, are endomorphisms of each module of this decomposition. Similarly, the basis of local, particle number conserving operators for the fermionic case is shown in Table~\ref{tab:model}. As a result, the dimension of the process tensor is reduced such that
\begin{equation}
\label{reduced_process_dimension}
    \dim[\mathcal{H}_{RB}] > \dim\left[\mathcal{H}_{RB}\left(N = L/2\right)\right] = {L \choose L/2} \times d_S^{n_B}.
\end{equation}

\begin{figure*}[t]
    \centering
\includegraphics[width=0.95\textwidth]{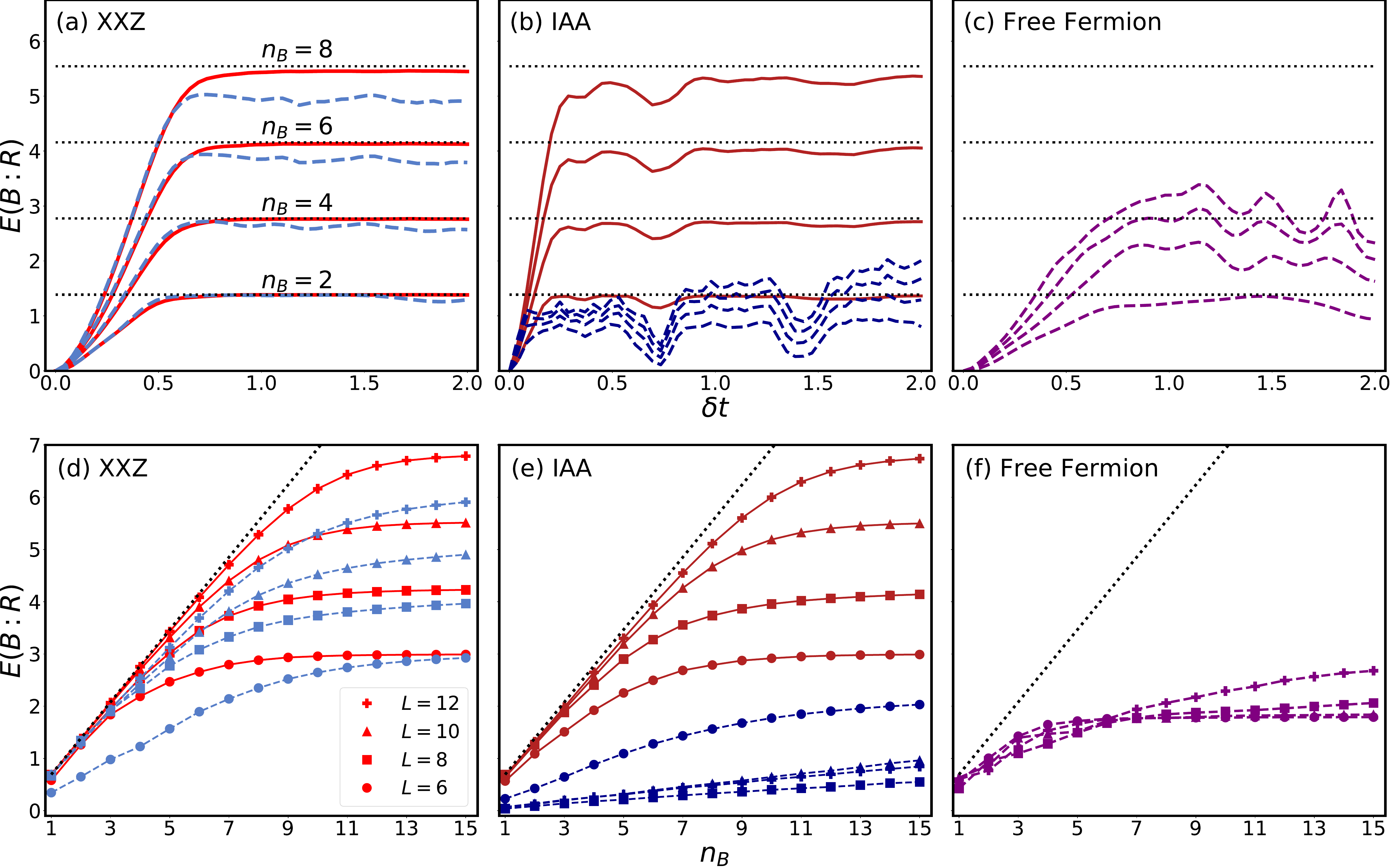}
    \caption{[(a), (b), (c)] Quantum dynamical entropy (QDE) as a function of the time between interventions, with system size $L = 14$ and increasing number of interventions, $n_{B} = 2, 4, 6, 8$. In (a) we plot results for the interacting integrable (dashed, blue) and chaotic (solid, red) XXZ model, together with the maximal value $\log(d_B)$ (dotted, black). In the chaotic case, the QDE saturates to its maximal value. In (b) we perform the calculation for the localized (dashed, blue) and chaotic (solid, red) IAA model. Similarly, we observe agreement between the chaotic and maximal dynamical entropy after sufficiently long times and see rapid saturation of the localized case below the maximal entropy. The free fermion case in (c) also saturates below this maximal entropy. [(d), (e), (f)] Scaling of QDE as a function of number of interventions, for fixed time between interventions given by $\delta t = 1.75$ and using the same models as upper panels. The black dotted line shows the Haar-random QDE estimated from Eq.~\eqref{eq:max_dyn_ent_scaling}. The chaotic cases show near-linear growth of dynamical entropy up to the maximal entropy for the finite-sized system $\log(d_R)$. The interacting integrable case in (e) has increasingly linear growth for larger system sizes, but saturates below the maximal entropy.  In all cases, as $L$ increases, the plots approach the maximal value.}
    \label{fig:dyn_ent}
\end{figure*}

Beyond using symmetry-preserving instruments, we can reduce the resources required to construct the dynamical and STE by considering only the reduced process on either the butterfly space $\ups_{B}$ or remainder space $\ups_{R}$. The reduced state on the butterfly space is given by
\begin{equation}
\label{eq:Upsilon_B}
    \ups_{B} = \sum_{\vec{x}, \vec{y}} \langle \ups_{R|\vec{y}}|\ups_{R|\vec{x}}\rangle |\vec{x}\rangle \langle \vec{y}|,
\end{equation}
and the reduced state on the remainder space is given by
\begin{equation}
\label{eq:Upsilon_R}
    \ups_{R} = \sum_{\vec{x}} |\ups_{R|\vec{x}}\rangle \langle \ups_{R|\vec{x}}|.
\end{equation}
Each state $|\ups_{R|\vec{x}}\rangle$ is defined by Eq.~\eqref{eq:output_state} and is constructed using numerically exact dynamics. By computing the output states for all combinations of multi-time interventions, the reduced process tensor can be constructed. Now that we have detailed the models, tools, and observables of interest, we can discuss the numerical results.

\section{Results}
\label{sec:results}

We discuss our numerical results for the figures of merit described in Sec.~\ref{sec:figs_of_merit}. We will analyze the dependence of these quantities on the Hamiltonians by studying their behavior with time between interventions $\delta t$, number of interventions $n_B$, and system size $L$. In general, we expect these quantities to depend on the initial state, so in all cases we consider the pure, low-entangled, non-stationary initial N\'eel state, shown in Eq.~\eqref{eq:neel}. Other initial states are considered in Appendix~\ref{app:initial_state_dep}, and we find qualitatively similar results. We perform time evolution using exact diagonalization with the help of the QuSpin library~\cite{10.21468/SciPostPhys.2.1.003, 10.21468/SciPostPhys.7.2.020} and all numerical results in this section use natural logarithms. We also report several analytical results.

\subsection{Quantum Dynamical Entropy}
\label{results:QDE}

We are interested in studying the quantum dynamical entropy (QDE), using its behavior for a random process as a benchmark. Ref.~\cite{FigueroaRomero_Modi_Pollock_2019} studied $n_B$-step processes generated from Haar-random unitary evolution and found that, as the dimension of the environment $d_E\to \infty$, the reduced process tensor $\Upsilon_B$ is maximally mixed. Here we consider a symmetry-restricted process tensor, given by a projection of Eq.~\eqref{pure_process_decomposition} onto the $\Delta N=0$ subspace. The corresponding reduced process tensor on the butterfly space is a maximally mixed state of dimension $d_S^{n_B}$ (c.f.~Eq.~\eqref{reduced_process_dimension}), with dynamical entropy 
\begin{equation}
\label{eq:max_dyn_ent_scaling}
    \langle E(B:R)\rangle_{\text{Haar}} = n_B \log (d_S).
\end{equation}
This result holds only for $d_S^{n_B} \ll d_R$. For instance, when $d_S^{n_B} \geq d_R$ the dynamical entropy is upper-bounded by the dimension of the remainder space as $E(B:R) \leq \log(d_R)$.

In Fig.~\ref{fig:dyn_ent}, we display our results for QDE computed using the many-body models and interventions shown in Table~\ref{tab:model}. The top row, Figs.~\hyperref[fig:dyn_ent]{2a}, \hyperref[fig:dyn_ent]{2b}, and \hyperref[fig:dyn_ent]{2c}, displays the dynamical entropy as a function of the time between interventions for different number of interventions $n_B$ and fixed system size, $L = 14$. In Fig.~\hyperref[fig:dyn_ent]{2a}, we compare the dynamical entropy for the interacting integrable and chaotic XXZ model. At short times between interventions, we observe similar behavior in both models. The QDE is zero for $\delta t = 0$, which is intuitive: no new information can be learned about the system by repeated probing if no evolution is allowed to occur between interventions. As $\delta t$ increases, we observe a growth in QDE as the subsystem has time to partially equilibrate between interventions. For large $\delta t$, we observe differences in the QDE for chaotic and non-chaotic dynamics, especially as the number of interventions $n_B$ grows large. The chaotic dynamics saturates to the maximum dynamical entropy for $n_B<L$, i.e.~$E(B:R) \approx \log(d_B) = n_B \log(d_S)$, indicated by the black dotted line. By contrast, the interacting integrable case does not saturate to this maximum.

In Fig.~\hyperref[fig:dyn_ent]{2b}, we show results for the IAA model in the chaotic and many-body localized phase. Here, the differences between these two models arise at much shorter timescales with the many-body localized case saturating at a much smaller value than the maximum. The chaotic case has fluctuations which are not observed in the chaotic XXZ case that are due to the microscopic details of the model; nevertheless, we see approximate saturation to the maximum dynamical entropy. Finally, the free fermion model in Fig.~\hyperref[fig:dyn_ent]{2c} does not saturate to the maximum allowed dynamical entropy. This large-$\delta t$ behaviour, like that of the integrable XXZ model in Fig.~\ref{fig:dyn_ent}(a), can be understood as a consequence of the constrained equilibration  of integrable systems due to conservation laws~\cite{rigol_thermalization_2008, Rigol2007, Cassidy2011, Calabrese2011}. We analyse the equilibration of these quantum processes in more detail in Sec.~\ref{sec:equilibration}.

The bottom row, Figs.~\hyperref[fig:dyn_ent]{2d}, \hyperref[fig:dyn_ent]{2e}, and \hyperref[fig:dyn_ent]{2f}, shows how QDE grows as a function of the total number of interventions $n_{B}$ for different system sizes $L = 6, 8, 10, 12$. We fix the time between interventions to be sufficiently large so we can distinguish chaotic behavior, $\delta t = 1.75$. Focusing on the XXZ model in Fig.~\hyperref[fig:dyn_ent]{2d}, we observe near-linear growth with $n_B$ in the chaotic case, agreeing closely with the Haar ensemble result from Eq.~\eqref{eq:max_dyn_ent_scaling}, which is shown by the black dotted line. As $n_B$ increases beyond the system size $L$, however, the QDE saturates to its maximum value $\log(d_R)$. Clearly, this saturation is a finite-size effect and we expect QDE for chaotic systems in the thermodynamic limit to grow indefinitely with $n_B$. The QDE for the interacting integrable model grows more slowly and also saturates at a point below the maximal entropy; however, as the system size increases the growth rate increases towards the Haar-random value. The IAA model in Fig.~\hyperref[fig:dyn_ent]{2e} displays similar features for the chaotic regime, but the many-body localized regime shows much smaller growth in the number of interventions. Despite this small growth rate, this model still appears to have linear scaling of dynamical entropy. Finally, in Fig.~\hyperref[fig:dyn_ent]{2f} the free fermion case does not agree with Eq.~\eqref{eq:max_dyn_ent_scaling} and saturates well below the maximal value. From these results, we can distinguish chaotic and non-chaotic dynamics in finite-sized systems based on their ability to maximize the dynamical entropy and their agreement with the Haar process after sufficiently long timescales.

One might wonder if QDE can unambiguously distinguish chaotic from non-chaotic dynamics as it does in the classical case, and if the signatures of quantum chaos are the same as classical chaos. This is in fact not the case. An analysis of the Alicki-Fannes dynamical entropy has been performed in Ref.~\cite{Cotler2018} for free fermions in the thermodynamic limit and indicates a maximal growth rate of the dynamical entropy---albeit requiring much larger timescales between interventions than in chaotic environments. Additionally, a class of models called Lindblad-Bernoulli shifts -- consisting of a circuit of SWAP gates with an infinite (large) environment -- are a highly regular type of dynamics, and yet maximize the QDE~\cite{Lindblad1979-tm,dowling_operational_2023}. Even from our finite-size study in Fig.~\hyperref[fig:dyn_ent]{2d}, we see the interacting integrable model approaching the Haar result as system size increases.

\subsubsection{Equilibration}
\label{sec:equilibration}



The results in Fig.~\ref{fig:dyn_ent}(a)-(c) indicate that the QDE equilibrates at large $\delta t$, with a saturation value and fluctuations that depend strongly on the nature of the model considered. In this subsection, we use existing results on the second law in isolated systems~\cite{PRXQuantum.6.010309} and the equilibration of process tensors~\cite{Dowling2023relaxationof,finitetime} to show that equilibration of the QDE occurs generically in many-body systems. In particular, we bound the probability that the QDE is outside a small distance from its equilibrium value---defined by the infinite-time average of the process. We use this to identify the effects of chaotic and non-chaotic dynamics on the equilibrium process tensor and the fluctuations around equilibrium. Ultimately, we show that equilibration of the QDE will occur when the number of interventions is small and the effective dimension of the initial state is large. Comparison of these results to our numerics indicates that the equilibrium QDE for non-chaotic models is less than in chaotic models.

Equilibration describes how an initially non-equilibrium state evolves in time to its equilibrium state. Understanding how and under what conditions equilibration occurs in many-body models has been a long-standing problem~\cite{Tasaki_1998, Popescu2006, Goldstein2006, Reimann_2008,ShortFinite, Malabarba2014, Styliaris2021, ShortSystemsAndSub}. In isolated, quantum systems, equilibration has been shown to occur in small, local subsystems due to the growth of entanglement, but requires that the spectrum of these models be non-degenerate and that the initial state comprises a large number of eigenstates, having a large effective dimension. Many studies of equilibration have focused on single-time states; however, the same techniques can be used to study equilibration in process tensors~\cite{Dowling2023relaxationof,finitetime}. Consider a process tensor, $\ups_{B}$, with $n_B$ number of interventions at times $t_1, ..., t_{n_B}$. The equilibrium process tensor is defined as the infinite-time average of the process, given by $\Omega_{B} = \langle\ups_{B}\rangle_{\infty}$, with time average defined as
\begin{equation}
    \langle \bullet \rangle_{\infty} = \prod_{i=1}^{n_B} \lim _{T_i \rightarrow \infty} \frac{1}{T_i} \int_0^{T_i} dt_i \left[ \bullet \right].
\end{equation}
Using results from the equilibration of single-time states~\cite{ShortSystemsAndSub}, it was shown that the time average of the trace distance between the process and its equilibrium is bounded as \cite{Dowling2023relaxationof, Wilde_2013},
\begin{equation}
\label{distance_bound}
    \left\langle D(\ups_{B}, \Omega_{B})\right\rangle_{\infty} \leq \frac{1}{2} M_{B} d^{n_B}_S \sqrt{\frac{2^{n_B} - 1}{d_{\text{eff}}(\rho)}}.
\end{equation}
Here, $D(\rho, \sigma) = \frac{1}{2}\|\rho - \sigma\|_1$ is the trace norm distance and $M_{B}$ is the combined total number of measurement outcomes on $B$, given by $M_{B} = d_S^{n_{B}}$. The effective dimension of the initial state, $\rho$, is defined as
\begin{equation}
    d_{\text{eff}}(\rho) = \frac{1}{\tr\left[\$(\rho)^2 \right]},
\end{equation}
where $\$(\rho) = \sum_{n}P_{n} \rho P_{n}$ is the overlap of $\rho$ and the eigenstates of the Hamiltonian, $P_{n} = |n\rangle \langle n|$. From Eq.~\eqref{distance_bound}, we have placed a bound on the time-averaged distance of the process from the equilibrium process.

Recently, it has been shown that the Shannon observational entropy of a local subsystem will generically increase in time in isolated systems due to equilibration~\cite{Linden2008, Safranek2019a,*Safranek2019b, PRXQuantum.6.010309}. In App.~\ref{Appendix:dyn_growth_t}, we apply the approach taken in  Ref.~\cite{PRXQuantum.6.010309} to show that on average for most times the quantum dynamical entropy will be equilibrated. We use the continuity of the R\'enyi-2 entropy~\cite{Chen_2016}, combined with Eq.~\eqref{distance_bound} to obtain the following bound on the time-averaged distance between the QDE and its equilibrium value,
\begin{equation}
\label{eq:dyn_ent_time_bound}
    \langle |S^{(2)}(\ups_{B}) - S^{(2)}(\Omega_{B})|\rangle_{\infty} \leq   2\langle D\rangle_{\infty} - \frac{d_{S}^{n_B}-2}{d_{S}^{n_B}-1} \langle D\rangle_{\infty}^2.
\end{equation}
We can then straightforwardly apply Markov's inequality to obtain a bound on the probability of the QDE being outside some distance from the QDE of the equilibrium process:
\begin{equation}
\label{eq:markov_bound}
\begin{aligned}
    &\mathds{P}\biggl[|S^{(2)}(\ups_{B}) - S^{(2)}(\Omega_{B})| \geq \biggr.
    \\& \qquad \qquad \biggl. \frac{1}{d_{\text{eff}}^{1/4}}\left(2-\frac{d_{S}^{n_B}-2}{d_{S}^{n_B}-1}\langle D \rangle_{\infty} \right)\biggr] \leq \frac{M_{B} d^{n_B}_S \sqrt{2^{n_B} - 1}}{2 d_{\text{\text{eff}}}(\rho)^{1/4}}.
\end{aligned}
\end{equation}

The bounds in Eqs.~\eqref{eq:dyn_ent_time_bound} and~\eqref{eq:markov_bound} will be satisfied for all dynamics provided the Hamiltonian has no degeneracies; however, the tightness of the bound is dependent on the number of interventions $n_B$ and the effective dimension of the initial state. The bound in Eq.~\eqref{eq:dyn_ent_time_bound} will be tight when
\begin{equation}
    d_{\text{eff}}(\rho)^{1/4} \gg \frac{1}{2} M_B d_S^{n_B} \sqrt{2^{n_B} - 1}.
\end{equation}
We have previously seen in Figs.~\hyperref[fig:dyn_ent]{2a},~\hyperref[fig:dyn_ent]{2b}, and~\hyperref[fig:dyn_ent]{2c}  that the QDE for the chaotic models increases as a function of $\delta t$ up to a saturation point with few fluctuations. The non-chaotic models reach some saturation point with smaller QDE and with larger fluctuations. The interacting integrable XXZ model and chaotic IAA models have similar-sized fluctuations. These results indicate that low-entangled states have smaller effective dimensions for non-chaotic models than for chaotic models. Our numerical results indicate that the equilibrium QDE for chaotic systems is described by the Haar average, suggesting a form of ergodicity in which $\langle E(B:R)\rangle_{\infty} \approx \langle E(B:R)\rangle_{\text{Haar}}$.

We emphasize that equilibration is not unique to chaotic dynamics as the result in Eq.~\eqref{eq:markov_bound} is independent of the unitary evolution. The true distinction between chaotic and non-chaotic equilibration of QDE is the strength of the fluctuations and the equilibrium value. We can use Figs.~\ref{fig:dyn_ent} to determine the equilibrium value for different types of dynamics and we observe that chaotic models agree with the Haar average and non-chaotic models have a smaller equilibrium QDE. It may be possible that both of these distinctions can be spoofed in integrable systems in the thermodynamic limit if the initial state has a sufficiently large effective dimension, therefore we do not view the process of equilibration as an unambiguous probe of chaotic dynamics.

\begin{figure}[t]
    \centering
    \includegraphics[width=0.45\textwidth]{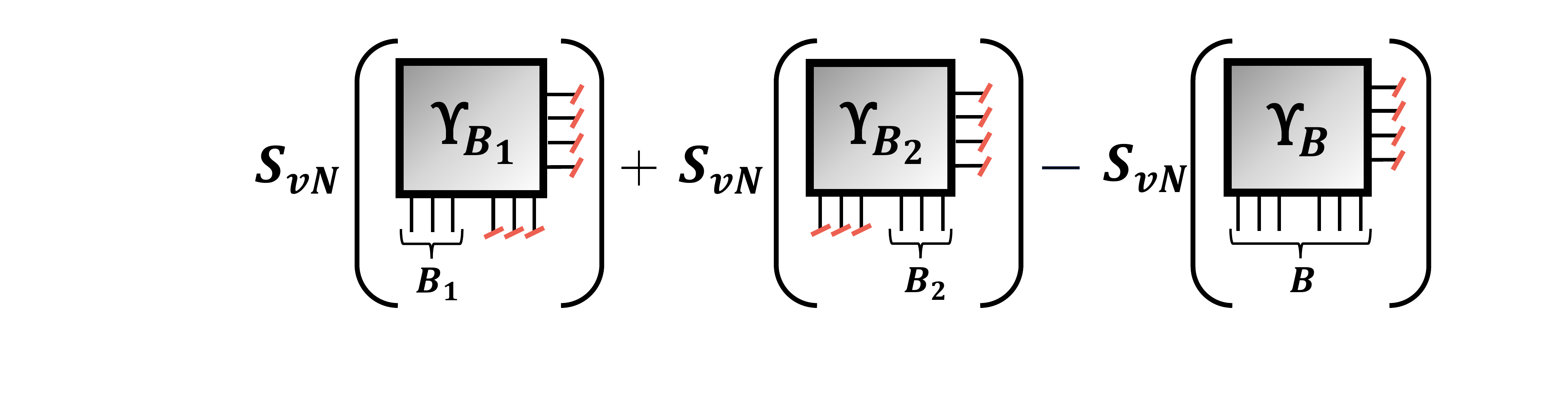}
    \includegraphics[width=0.43\textwidth]{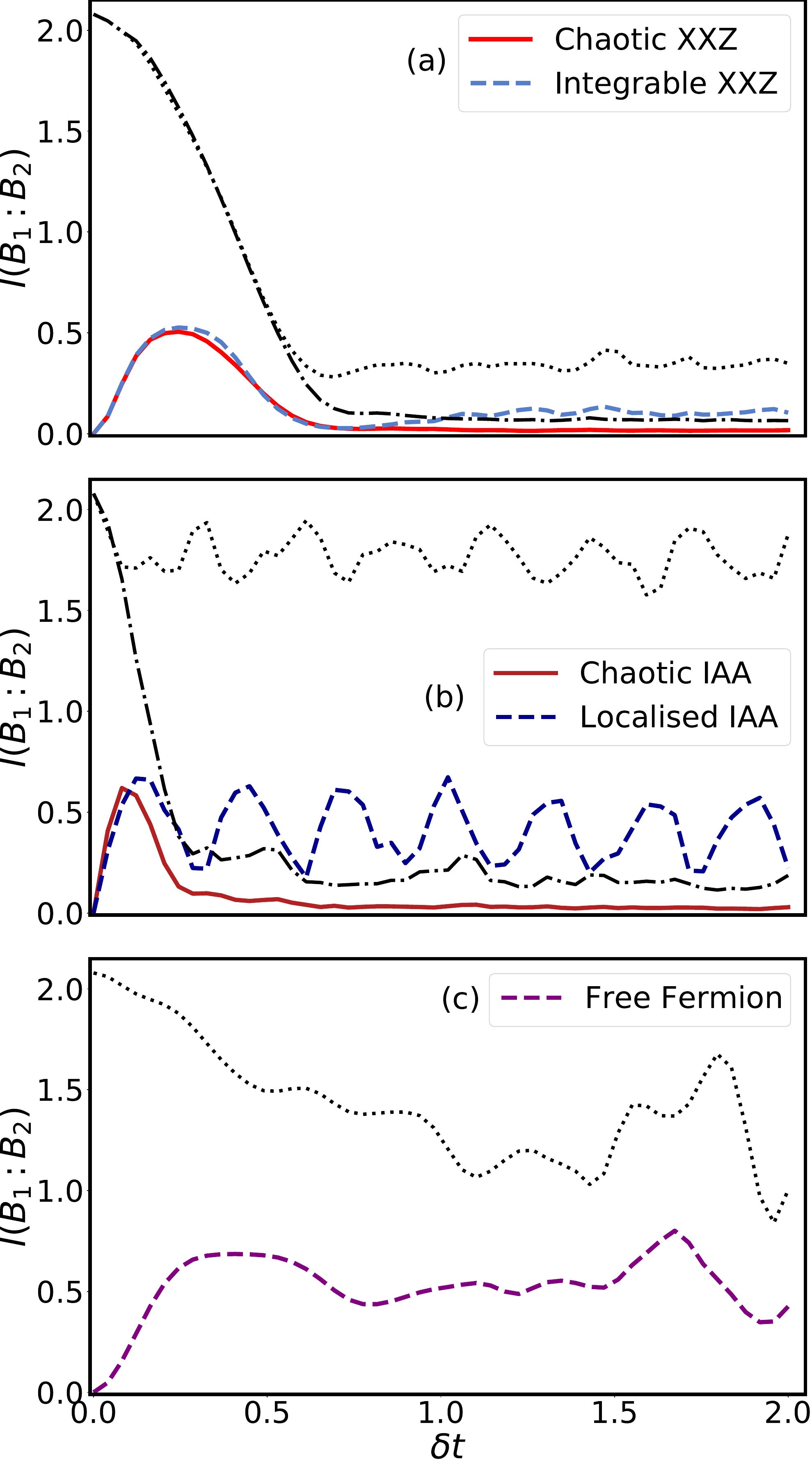}
    \caption{Temporal mutual information between $B_1$ and $B_2$. We consider system size $L = 12$, number of interventions, $n_{B} = 6$ with $n_{B_1}, n_{B_{2}} = 3$, and state initially prepared in the Neel state from Eq.~\eqref{eq:neel}. (a) Chaotic (solid, red) and interacting integrable (dashed, blue) XXZ model with bound from Eq.~\eqref{eq:Markovian_quantifier} given by the dashed-dotted line and dotted line for the chaotic and integrable case, respectively. (b) Chaotic (solid, red) and localized (dashed, blue) IAA model with bound also given by dashed-dotted and dotted black lines. (c) Free fermion (dashed, purple) model with bound given by dotted line.}
    \label{fig:mut_inf}
\end{figure}

\subsubsection{Temporal Correlations and Markovianization}

By comparing with the Haar average process, we have interpreted the QDE as a measure of the randomness of the dynamics. The behavior of Haar random process tensors have previously been studied and is shown to be typically Markovian for large environment dimensions~\cite{Pedro2020}. Additionally, chaotic dynamics have been studied as a sufficient mechanism for producing Markovian behavior~\cite{Strasberg_2023, odonovan2024}. It is thus natural to probe non-Markovian correlations in chaotic, short-ranged lattice models to see if they are sufficiently random to produce Markovian behavior in the process tensor.

To study this problem, we first define a quantifier of temporal correlations. Previously, the mutual information been established as a quantifier of non-Markovian correlations~\cite{processtensor, processtensor2}: those that are transmitted through the environment only. Here, we use the mutual information between temporal regions $B_1$ and $B_2$ of the process tensor, defined as 
\begin{equation}
\label{eq:mut_inf}
    I(B_1:B_2) = S_{\text{vN}}(\ups_{B_1}) +  S_{\text{vN}}(\ups_{B_2}) -  S_{\text{vN}}(\ups_{B})
\end{equation}
with $\ups_{B_{2(1)}} = \tr_{B_{1(2)}}(\ups_{B})$ and $S_{\text{vN}}$ the Von-Neumann entropy. We previously defined the QDE using the Renyi-2 entropy as it was more efficient numerically, allowing us to obtain larger system sizes and time steps. However, the Renyi-2 entropy does not obey subadditivity so the Renyi mutual information is not operationally well-defined and can be negative~\cite{Kormos_2017}. For this section, we employ the Von Neumann entropy to get a well-defined probe of temporal correlations. Note, since we do not employ a so-called causal break~\cite{milz_quantum_2021}, Eq.~\eqref{eq:mut_inf} quantifies \textit{all} temporal correlations---those which pass through both system and the environment---and thus upper-bounds the non-Markovian correlations.

We show in App.~\ref{Appendix:Dyn_Ent_markov} that the growth of dynamical entropy with $n_B$ bounds temporal correlations as quantified by the mutual information,
\begin{equation}
\label{eq:Markovian_quantifier}
    I(B_1: B_2) \leq n_{B_2}\left[\log(d_S) - \frac{S_{\text{vN}}(\ups_{B}) - S_{\text{vN}}(\ups_{B_1})}{n_{B} - n_{B_1}}\right].
\end{equation}
Here, $n_B = n_{B_1} + n_{B_2}$ is the total number of interventions which is the sum of the number of interventions made in $B_1$ and $B_2$. By non-negativity of the mutual information, this bound is tight when the growth rate of the QDE is maximal,
\begin{equation}
    \frac{S_{\text{vN}}(\ups_{B}) - S_{\text{vN}}(\ups_{B_1})}{n_{B} - n_{B_1}} = \log(d_S),
\end{equation}
in which case the mutual information is zero, $I(B_1:B_2) = 0$. This result directly links the dynamical entropy, a probe of randomness, to the decay of temporal correlations in time. As discussed in Sec.~\ref{sec:PPE}, the dynamical entropy is a probe of the first moment of the PPE, indicating that only first-order randomness is sufficient to achieve Markovianity. Note, that the maximal growth rate of dynamical entropy with $n_B$ is sufficient but not necessary for Markovianity; regular, non-chaotic dynamics can also lead to Markovian behavior~\cite{PhysRevLett.133.170402}.

\begin{figure*}[t]
    \centering
    \includegraphics[width=0.95\textwidth]{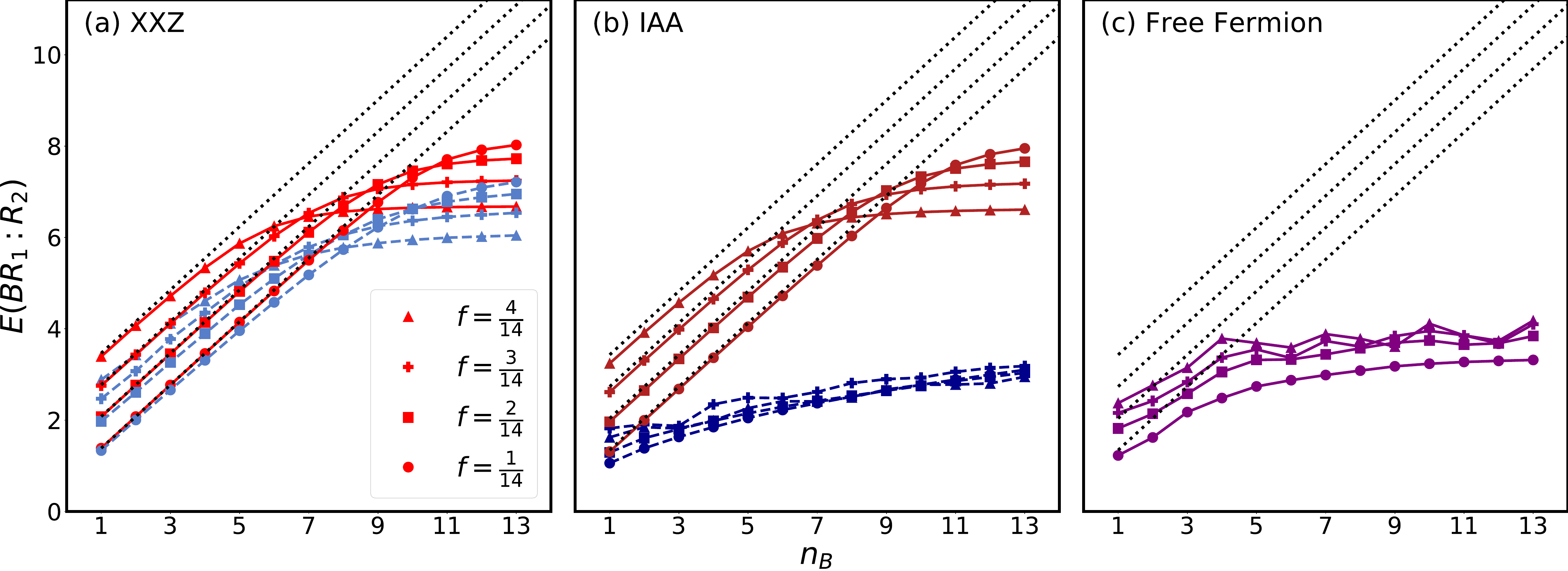}
    \caption{Spatiotemporal entanglement (STE) for increasing number of interventions with time between interventions $\delta t = 1.75$ and total system size $L = 14$. We consider different fractions of the output subspace traced out, $f = \frac{L_{R_1}}{L}$. In (a), we plot the results for chaotic (red, solid) and interacting integrable (blue, dashed) XXZ models observing initial linear scaling with $n_B$ for both cases with a smaller overall STE for the interacting integrable case, disagreeing with the Haar average result. We also observe substantial disagreement between the non-chaotic case and the Haar average result from Eq.~\eqref{eq:Haar_average_ST} (black, dotted) for larger fractions $f$ due to finite-size effects. (b) displays the results for the chaotic (red, solid) and localized (blue, dashed) phases of the IAA model. The chaotic IAA has a similar behavior as the chaotic XXZ and the many-body localized case has a much smaller growth rate. The free fermion, in (c), also disagrees substantially with the Haar average.}
    \label{fig:st_ent}
\end{figure*}

In Fig.~\ref{fig:mut_inf}, we plot the mutual information between parts of the butterfly space as a function of time. We take $L = 12$ and consider $n_B = 6$ with $n_{B_1}, n_{B_2} = 3$. In Fig.~\hyperref[fig:mut_inf]{3a}, we compute the mutual information for the chaotic and interacting integrable XXZ models. We see similar initial growth of non-Markovian correlations at short times. After sufficiently long timescales, these correlations decrease in the chaotic case and the bound in Eq.~\eqref{eq:Markovian_quantifier}, given by the black, dashed-dotted line, becomes tight. The interacting integrable case also has small non-Markovian correlations but with larger fluctuations and a less tight bound, given by the black, dotted line. In the IAA model, Fig.~\hyperref[fig:mut_inf]{3b}, we see that the chaotic case has similar behavior to the chaotic XXZ model with rapid growth and decay of the non-Markovian correlations, and with a tight bound. The many-body localized phase displays distinct behavior having long-lived oscillating non-Markovian correlations and an extremely loose bound. Similarly, the free fermion case, Fig.~\hyperref[fig:mut_inf]{3c}, also displays long-lived non-Markovian correlations. In the thermodynamic limit and with a sufficiently long time between interventions, it is likely that most models will have near-Markovian behavior, as information becomes hidden in the environment.

From the results presented above, Markovianity and memorylessness appears to generically occur in chaotic systems after equilibration timescales. As previously discussed, Markovianity is not a unique features of chaotic dynamics and may occur in non-chaotic systems. This result appears to contradict the previously studied behavior of another probe of temporal correlations called the temporal entanglement~\cite{sim_complex_Filippov_19,PhysRevResearch.6.033021, boucomas2024, lerose_overcoming_2023}. This quantity is associated with the complexity of process tensor methods used for simulating local dynamics interacting with complex environments. It has been shown that temporal entanglement has distinctive scaling behavior in time for free, interacting integrable, and chaotic models: area-law~\cite{Lerose2021},  logarithmic~\cite{Giudice2022}, and volume-law~\cite{PhysRevX.13.041008}, respectively. Here, our results indicate that chaotic models appear Markovian, whereas temporal entanglement shows extensive growth for chaotic models. We attribute this difference to the large time $\delta t$ between interventions that we consider, as compared to the small or infinitesimal timestep of Trotterised or Floquet circuit time evolution studied in previous works~\cite{sim_complex_Filippov_19,Lerose2021,Giudice2022, lerose_overcoming_2023,PhysRevX.13.041008,PhysRevResearch.6.033021,boucomas2024}. This is consistent with the expectation that highly complex time evolution on short timescales is associated with emergent Markovianity on longer timescales~\cite{odonovan2024,Strasberg_2023}. However, we leave a careful investigation of this connection to future work.

In the next section, we will numerically compute the spatiotemporal entanglement, an extension to the quantum dynamical entropy, which is expected to be a more refined probe of chaos in many-body systems and is known to be able to distinguish Lindblad-Bernoulli dynamics from chaotic dynamics~\cite{dowling_operational_2023}.

\subsection{Spatiotemporal Entanglement}
\label{results:STE}

We now numerically analyze the spatiotemporal entanglement (STE) and are interested in whether this quantifier of entanglement within the process tensor can faithfully distinguish between chaotic and non-chaotic models, given that it contains additional information about the complexity of the output state and about information scrambling.

We will compare the STE across a bipartition $BR_1:R_2$ to the STE for a Haar averaged spatiotemporal state, $\langle \Upsilon_{BR_1}\rangle_{\text{Haar}} = \frac{1}{d_R} \mathds{1}\otimes (\frac{1}{d_{S}})^n\mathds{1}_{B}$, as derived in App.~\ref{Appendix:st_ent_haar}. We them approximate the Haar average STE as
\begin{equation}
\label{eq:Haar_average_ST}
    \langle E(BR_1:R_2)\rangle_{\text{Haar}} \approx n_B \log (d_S) + \log(d_{R_1}).
\end{equation}
This result will generally neither lower- nor upper-bound the true value of the Haar-averaged STE; regardless, we use this to compare with our numerical calculations and show that it closely approximates the behavior of chaotic dynamics. Here, we will have $d_{S} = 2$ which is the dimension of the subspace on which our interventions act and $d_{R_1} = 2^{L_{R_1}}$ is the dimension of the subspace $R_1$.

Our numerical results for the STE are shown in Fig.~\ref{fig:st_ent}. Here, we consider bipartitions $BR_1:R_2$ with fraction $f = L_{R_1}/L$ of the remainder space included in the bipartition. We plot this quantity as a function of the number of interventions for a variety of fractions $f$ with fixed time between interventions given by $\delta t = 1.75$ and a total system size of $L = 14$. In Fig.~\hyperref[fig:st_ent]{4a}, we plot the STE for the chaotic and interacting integrable XXZ model. We observe approximately linear scaling of this entropy and agreement with the Haar process, represented by the black, dotted line, for small $n_{B}$ and $f$ for both models. As both $n_B$ and $f$ increase, the two models begin to substantially disagree with the Haar average. Unlike the dynamical entropy, we are able to observe differences between the chaotic and interacting integrable model for a small number of interventions, provided that a larger fraction of the $R_1$ is included in the bipartition. In Fig.~\hyperref[fig:st_ent]{4b}, we perform the same calculation for the IAA model in the chaotic and many-body localized phase. We observe suppression of the growth of the STE in the many-body localized phase due reduced spatial entanglement growth which is known to occur in these models~\cite{PhysRevLett.110.260601, PhysRevB.90.174202, PhysRevB.77.064426, PhysRevLett.109.017202, Lukin_2019}. In contrast, the chaotic IAA model has linear growth in $n_{B}$, agreeing with the Haar averaged result more closely. Finally in the free fermion case, in Fig.~\hyperref[fig:st_ent]{4c}, we observe sublinear growth in $n_{B}$.

From these results, we observe that the inclusion of a larger portion of the output state enables us to distinguish chaotic and non-chaotic dynamics for smaller values of $n_B$. This difference is due the growth of entanglement between parts of the remainder space which, in finite-sized systems, are known to display distinct behavior~\cite{De_Chiara_2006, PhysRevA.78.010306, PhysRevLett.111.127205, PhysRevLett.110.260601, Alba_2018}. The STE does not have distinctive scaling behavior with $n_B$ so it does not improve substantially upon the results from the QDE. However, we have only considered a subset of possible bipartitions of the process. It is conceivable that the minimum STE over all possible bipartitions would have different scaling behaviors; however, this minimization is onerous to perform. Noticing the insufficiencies of QDE and STE, and that inclusion of output-state complexity can improve the distinguishability of chaotic and non-chaotic dynamics, we are motivated to probe the output state complexity further by studying higher-order moments of the PPE.

\subsection{Projected Process Ensemble}
\label{results:HOM}

\begin{figure*}[t]
    \centering
    \includegraphics[width=1\textwidth]{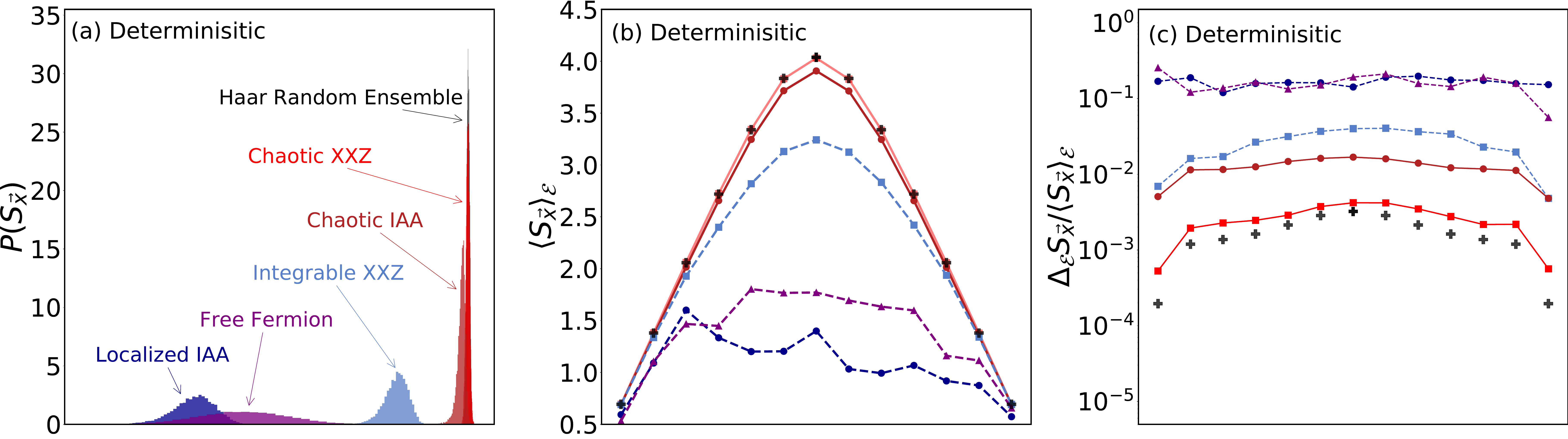}
    \includegraphics[width=1\textwidth]{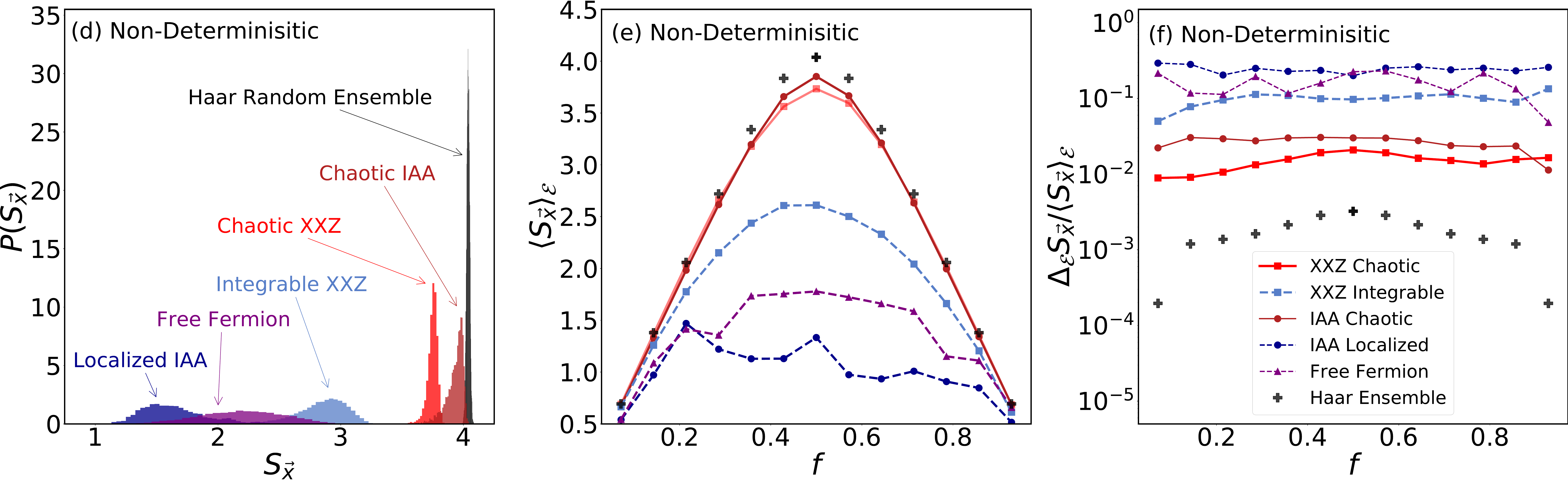}
    \caption{Results for the bipartite entanglement entropy of the PPE using deterministic interventions given by $\{\mathds{1}_1, \sigma^z_{1}\}$ shown in [(a), (b), (c)] and non-deterministic interventions given by $\{|0\rangle \langle 0|_1, |1\rangle \langle 1|_1\}$ shown in [(d), (e), (f)]. We compute the entanglement between different bipartitions for a fraction, $f = \frac{L_{R_1}}{L}$, traced out. (a) Distribution of the bipartite entanglement entropy distribution of the PPE for $f = \frac{1}{2}$, system size $L = 14$, number of interventions $n_{B} = 14$ and time between interventions $\delta t = 1.75$. For the XXZ model we see that the chaotic case has a larger mean and smaller variance than the integrable model. This behavior is also observed for the chaotic and localized IAA model as well as the free fermions case. (b) Mean bipartite entanglement entropy of the PPE ensemble as a function of the fraction of of the subspace, $f$. The chaotic and interacting integrable XXZ models saturate to a mean below the value of the Haar mean. The interacting integrable model agrees with the Page value for small and large fraction $f$, but saturates much lower than the chaotic models as $f = \frac{1}{2}$. The localized IAA and the free fermion model disagree substantially from the Haar mean for all $f$. (c) We observe similar behavour of the variance with the chaotic cases agreeing more closely with the Haar ensemble. [(d), (e), (f)] We repeat the above calculations using non-deterministic interventions. We observe that the action of projections at multiple times reduces the entanglement of all models, but still enables us to distinguish chaotic and non-chaotic dynamics.}
    \label{fig:PPE_dist}
\end{figure*}

In this section, we numerically compute the higher moments of the projected process ensemble (PPE). In particular, we compute the mean, $\langle S_{\vec{x}}\rangle_{\mathcal{E}}$, and variance, $\Delta_{\mathcal{E}} S_{\vec{x}}$, of the bipartite entanglement entropy of the output states of the process tensor for a fixed basis of measurements defined in Sec.~\ref{sec:PPE}. These probes are expected to be more capable of distinguishing chaotic and non-chaotic dynamics as is seen in deep thermalization. Unlike the QDE and STE, the higher PPE moments depend on the basis of interventions. We compare the PPE constructed using deterministic and non-deterministic interventions. The deterministic interventions are given by $\tfrac{1}{\sqrt{2}}\left\{\mathds{1}_1, \sigma^z_{1}\right\}$ and $\tfrac{1}{\sqrt{2}}\bigl\{c^{\dagger}_1 c_1 + c_1 c^{\dagger}_1, c_1^{\dagger}c_1 - c_{1}c^{\dagger}_{1}\bigr\}$ for the spin and fermionic models, and the non-deterministic interventions are given by $\{|0\rangle \langle 0|_1, |1\rangle \langle 1|_1\}$ and $\{c^{\dagger}_{1}c_1, c_{1}c_{1}^{\dagger}\}$. Both of these types interventions preserve the particle number symmetry.

As we have done in the previous sections, we are interested in comparing our numerical results to the Haar averaged process. In App.~\ref{Appendix:PPE_moments}, using methods from the theory of symmetric subspaces~\cite{Church_of_sym, PRXQuantum.4.010311}, we show that the $k$-th order moment of PPE with independently averaged Haar-random unitary dynamics is given by
\begin{equation}
\label{eq:Haar_moment}
\begin{aligned}
&\langle \ups^{(k)}_{R}\rangle_{\text{Haar}} = \int_{U_1}...\int_{U_n}\left[\ups_{R}^{(k)}[U_{1:n+1}]\right] 
\\& \qquad \qquad \; \; = \frac{\sum_{\pi \in S_{k}}P_{\pi}}{d_{R}(d_{R}+1)...(d_{R}+k-1)}.
\end{aligned}
\end{equation}
Here, $P_{\pi}$ is a representation of the symmetric group, $S_{k}$, whose elements are given by
\begin{equation}
    P_{\pi} = |i_{\pi^{-1}(1)} i_{\pi^{-1}(2)} ... i_{\pi^{-1}(k)}\rangle \langle i_1 i_2 ...i_{k}|,
\end{equation}
with $\pi \in S_{k}$. These moments are identical to the moments of the Haar ensemble and are independent of the choice of intervention operator or the number of interventions. In general, process tensors have more structure compared to pure states due to causality constraints, so we might expect different entanglement structure to arize; however, the action of a fixed basis of measurements breaks these constraints and the independent Haar averaging washes out the features associated with this constraint, leaving us with an ensemble of randomly sampled states. We will use this as a benchmark to compare with the PPE results for the Hamiltonian systems shown in Table~\ref{tab:model}.

Methods to compute the moments of bipartite entanglement entropies from Haar ensembles have been developed with the Bianchi-Dona distribution~\cite{Bianchi_Hackl_Kieburg_Rigol_Vidmar_2022, PhysRevD.100.105010, Ent_Haar_2016, Ent_Variance_2017, PhysRevD.100.105010} being most relevant as it applies to randomly sampled states from a Hilbert space with a conserved quantity $N$, having the form 
\begin{equation}
\label{eq:hilb_space_center}
    \mathcal{H}_R(N) = \bigoplus^{N}_{n=0} \mathcal{H}_{R_2}(n)\otimes \mathcal{H}_{R_1}(N - n).
\end{equation}
In our case, $N$ will correspond to the particle number and $n_{j}$ will be the particle number within $\mathcal{H}_{R_{2}}$. However, results from these works cannot be applied to the R\'enyi-2 entropy due to the non-linearity of the logarithm. Instead, we numerically compute the mean and variance of the bipartite R\'enyi-2 entanglement entropy by randomly sampling state from the Hilbert space in Eq.~\eqref{eq:hilb_space_center}.

We first look at the distribution of the bipartite entanglement entropy of the output states, shown in Figs.~\hyperref[fig:PPE_dist]{5a} and~\hyperref[fig:PPE_dist]{5d}, for the deterministic and non-deterministic interventions defined in Sec.~\ref{sec:pure_pt}. We generate these distributions for system size $L = 14$ and fixed time between interventions $\delta t = 1.75$. We considered $n_{B} = 14$ interventions so the number of states in the PPE is $2^{14}$. We compute the R\'enyi-2 entanglement entropy of a reduced state on $R_{1}$ with $f = L_{R_1}/L = 1/2$. We then construct the distribution by selecting small windows of entropy and counting the number of states whose entanglement entropy are within these windows. As a reference, we also plot the distribution of the bipartite entanglement for an ensemble of $2^{14}$ randomly sampled pure states from $\mathcal{H}_R$. In all cases, we normalize these distributions. 

In the deterministic case, Fig.~\hyperref[fig:PPE_dist]{5a}, the entanglement entropy distribution for the chaotic XXZ and IAA models appear to closely agree, in both mean and variance, with the distribution computed using Haar randomly sampled states. Conversely, all non-chaotic models show distinct differences with the Haar ensemble distribution. The interacting integrable model appears to be most similar in both mean and variance. The many-body localized model has a smaller mean and variance than the free fermion model. In the non-deterministic case, Fig.~\hyperref[fig:PPE_dist]{5d}, we still observe the chaotic models agreeing more closely with the Haar ensemble. However, the disentangling effect of the projection operators appears to reduce the ergodic behavior of these models such that we observe less agreement in both the mean and variance of the distributions. 

We now look at the profile of the mean and the standard deviation (in units of the mean). First, we focus on the deterministic case, shown in Figs.~\hyperref[fig:PPE_dist]{5b} and \hyperref[fig:PPE_dist]{5c} with the mean and standard deviation plotted as a function of the fraction of the subspace traced out $f = L_{R_1}/L$. In Fig.~\hyperref[fig:PPE_dist]{5b}, we observe that the chaotic and non-chaotic models disagree most strongly near $f = 1/2$. The chaotic XXZ and IAA models both agree more closely with the Haar ensemble result than the non-chaotic models. The interacting integrable XXZ model, which is known to be sensitive to perturbations, has larger mean than both the non-interacting and many-body localized models, which are known to be robustly non-ergodic. In Fig.~\hyperref[fig:PPE_dist]{5c}, we plot the coefficient of variation of the distribution---defined as the standard deviation divided by the mean. We see that the chaotic XXZ model agrees more closely with the Haar average result than the chaotic IAA model and the non-chaotic models are farther from the Haar results as we saw in Fig.~\hyperref[fig:PPE_dist]{5b}. We also observe the non-interacting and many-body localized models exhibiting robust non-chaotic behavior in the presence of these interventions, in contrast to the interacting integrable model which has a narrower standard deviation. 

The mean and standard deviation for non-deterministic interventions are shown in Figs.~\hyperref[fig:PPE_dist]{5e} and \hyperref[fig:PPE_dist]{5f}. We observe an overall reduction of the mean entanglement entropy compared to the deterministic interventions. The impact of the action of the non-deterministic interventions appears more prominent in the chaotic and interacting integrable XXZ models compared to the chaotic IAA model which now has a larger mean entanglement entropy than the chaotic XXZ model. The many-body localized and non-interacting models both have reduced mean entanglement entropy. The standard deviation increases for all models when using projective operators resulting in less agreement with the Haar ensemble result. These results are consistent with results from the study of monitored many-body dynamics and measurement-induced phase transitions~\cite{PhysRevB.99.224307, Skinner2019, PhysRevB.98.205136, PhysRevB.100.134306}. In particular, the action of measurements throughout the evolution has a disentangling effect which is typically observed to be much stronger in non-chaotic models, and is discussed further in Sec.~\ref{sec:discussion}.

\begin{figure}
    \centering
    \includegraphics[width=0.48\textwidth]{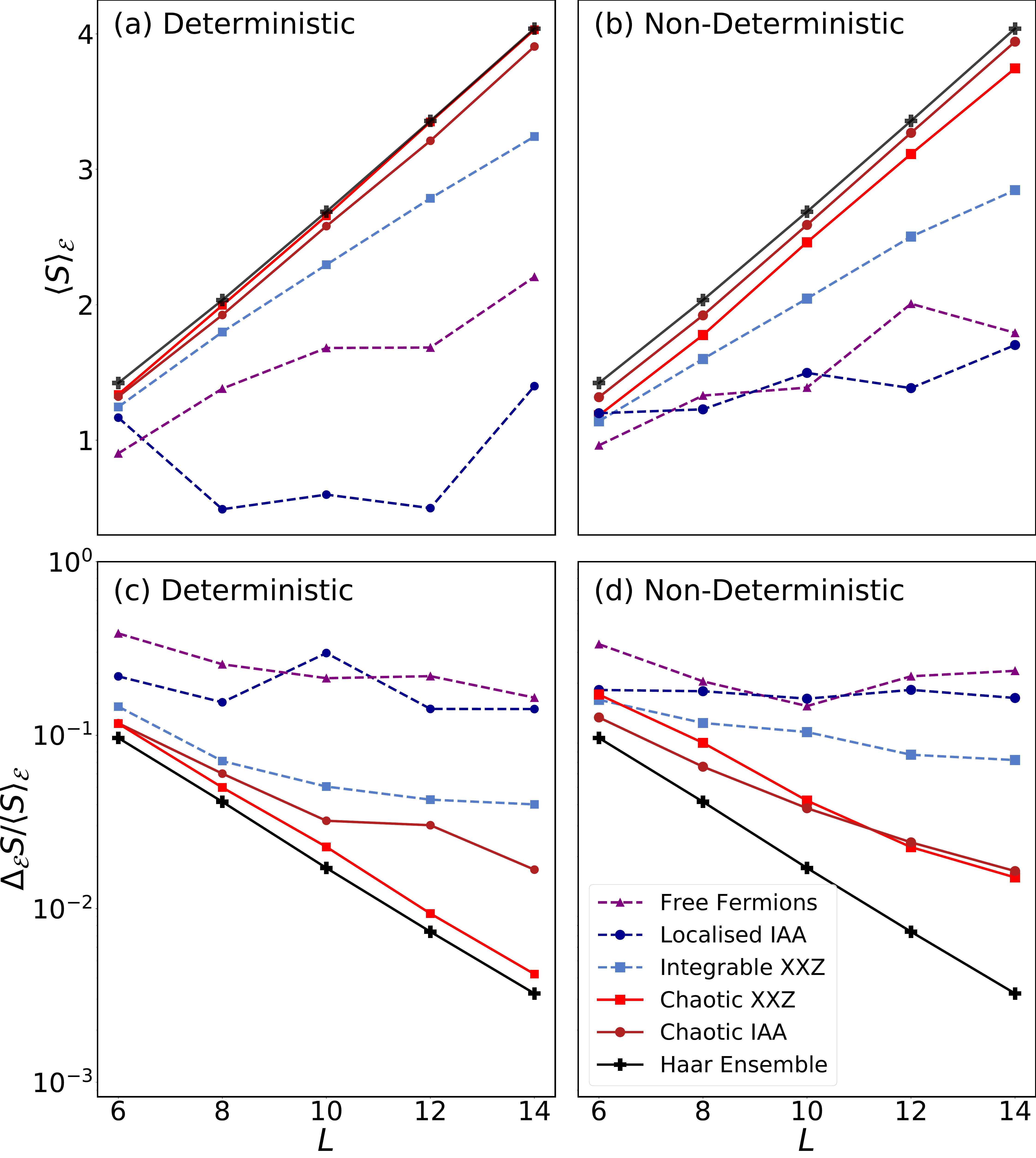}
    \caption{System size scaling of the mean [(a), (b)] and standard deviation [(c), (d)] of the bipartite entanglement distribution of the PPE for $f = \frac{1}{2}$, $n_B=14$, and fixed time between interventions $\delta t = 1.75$. We consider deterministic and non-deterministic interventions shown in Tab.~\ref{tab:model}.}
    \label{fig:PPE_t_L}
\end{figure}

To further characterize the effect of deterministic and non-deterministic interventions, we study the mean of the entanglement entropy as a function of system size. These calculations are performed for $n_B = 14$ and $\delta t = 1.75$. In Figs.~\hyperref[fig:PPE_t_L]{6a}, we analyze the system size scaling of the mean bipartite entanglement entropy for the deterministic operators. We observe linear growth of the mean in system size for the chaotic XXZ and IAA models, agreeing with the behavior of the Haar ensemble. The interacting integrable XXZ model also appears to display linear growth in system size, but with a smaller slope. The mean entropy of the free fermion model grows with system size, and the many-body localized IAA model does not, indicating that many-body localization is more robust against interventions. In Fig.~\hyperref[fig:PPE_t_L]{6b}, we consider non-deterministic interventions, where the mean entanglement of the PPE is smaller compared to the deterministic case. Nevertheless, the mean entanglement of chaotic models still displays linear growth in system size. Again, we see approximate linear growth in the interacting integrable XXZ model, but we see what appears to be approximate linear behavior in the many-body localized and free fermion model.

The results for the chaotic models indicate that we are within the entangling phase where information scrambling dominates. The chaotic XXZ model is more strongly affected by the use of projective measurements than the chaotic IAA model. This may be an indication of distinct scrambling rates in these models, but we leave a detailed analysis of this for future work. Due to numerical constraints, it is difficult to obtain system sizes larger than what is considered here. As a result, it is difficult to make strong claims about the behavior of the mean entropy in the thermodynamic limit. Additionally, we have not analyzed the behavior of the mean as a function of time between interventions. The discrepancy between the chaotic and non-chaotic models in this case may be a result of finite time and finite-size effects. It has been shown that even in non-interacting free fermion models, the entanglement entropy grows extensively after sufficiently long times~\cite{Alba_2018}. This suggests that extensive growth of the mean entanglement entropy may not be unique to chaotic systems; however, the timescale at which extensive behavior occurs and the rate of increase with system size may depend on whether the dynamics is chaotic, which can be observed in Fig.~\hyperref[fig:PPE_t_L]{6}.

We now look at the system-size scaling of the standard deviation of the entanglement distribution to see if this is more capable of probing chaos and to observe the impact of different types of interventions. First, we consider deterministic interventions in Fig.~\hyperref[fig:PPE_t_L]{6c}. Here, the chaotic XXZ model is observed to have an exponentially decreasing standard deviation in system size, qualitatively agreeing with the Haar ensemble. The chaotic IAA model has a larger standard deviation than the chaotic XXZ model and has less clear exponential decay. The interacting integrable model has still larger standard deviation and also does not exhibit exponential decay with system size. Finally, the localized IAA and free fermion models have much larger standard deviation and are not well described by the Haar ensemble. Non-deterministic interventions, see Fig.~\hyperref[fig:PPE_t_L]{6d}, have larger overall standard deviation compared to the deterministic case. The standard deviation of the chaotic XXZ model is larger and the rate of exponential decay in system size is smaller. Again, we see that the chaotic IAA model appears to be less affected by the use of non-deterministic interventions.

From this finite-size scaling analysis, the higher-order moments of the entanglement entropy distribution of the PPE appear more capable of distinguishing chaotic and non-chaotic dynamics, with exponential decay of the coefficient of variation associated with chaotic dynamics. We find that more robust forms of ergodicity-breaking, such as non-interacting models and many-body localization, are more easily distinguished from chaos than sensitive mechanisms of non-ergodicity, such as interacting integrable models. We also observe more clear distinctions when using deterministic interventions as opposed to non-deterministic interventions which disentangle the output states and introduce non-ergodic behavior.

\section{Discussion and Conclusion}
\label{sec:discussion}

In this work, we investigated the entanglement properties of the pure process tensor in a variety of quantum many-body models including those which exhibit chaotic behavior and which have distinct ergodicity-breaking mechanisms: interacting integrable, non-interacting, and many-body localized. The first of these entanglement properties, the quantum dynamical entropy (QDE) and spatiotemporal entanglement (STE), were developed in Ref.~\cite{dowling_operational_2023} as part of an operational definition of the butterfly effect in quantum systems. These quantities displayed distinct signatures of chaotic behavior; namely, near maximal growth of QDE with the number of interventions and larger total STE. However, due to known deficiencies with these measures, we extended our analysis to higher-order probes of chaotic dynamics. We defined an ensemble of pure states constructed by acting a fixed basis of local interventions at multiple times throughout the evolution---called the projected process ensemble (PPE)---which is analogous to state ensembles constructed in the study of deep thermalization~\cite{Ippoliti2022solvablemodelofdeep}. We focus on the properties of the mean and variance of the bipartite entanglement entropy, finding that the system size scaling of the variance---when using deterministic interventions---has distinct behavior for chaotic and non-chaotic dynamics. We have therefore established the PPE as a tool to characterize chaos and spatiotemporal complexity in quantum stochastic processes.

The QDE used in this paper has a number of subtle differences to the Alicki-Fannes dynamical entropy~\cite{Alicki1994-dc}, which may have implications in the classical limit. The first difference is the use of the R\'enyi-2 entropy, rather than the Von Neumann entropy, to quantify entanglement. Qualitatively, this difference does not impact our conclusions about the behavior of dynamical entropy in the quantum regime. In the classical limit, however, this choice may be important as the Alicki-Fannes entropy is expected to become equivalent to the classical dynamical entropy---uniquely defined as the Shannon entropy of the multi-time phase space distribution. A potential direction for future work would be to use the classification of classical process tensors~\cite{Taranto2024characterising, Milz2020prx, strasberg2019} to understand the behavior of the R\'enyi-2 QDE in the classical limit. The second difference is the use of local operators which act on a single site of the many-body model. It is interesting to speculate as to whether a choice of non-local interventions can more easily distinguish chaotic and non-chaotic dynamics; however, we leave this for future work.

We have studied temporal correlations for a variety of different models. In particular, we showed that the growth of QDE with the number of interventions bounds correlations in time. The linear growth of QDE with near-maximal rate---which is seen for chaotic dynamics---results in almost Markovian behavior. This result is consistent with previous results studying Haar-random processes~\cite{FigueroaRomero_Modi_Pollock_2019} and chaotic environments in open quantum systems~\cite{odonovan2024}. In the future, we are interested in investigating how Markovianity at the level of the process tensor impacts the growth of temporal entanglement. Temporal entanglement is a measure of simulation complexity~\cite{sim_complex_Filippov_19,PhysRevResearch.6.033021, boucomas2024, lerose_overcoming_2023} that also quantifies temporal correlations in multi-time states, but behaves differently to the mutual information. Temporal entanglement has been shown to have distinctive scaling behavior in time for chaotic and non-chaotic dynamics~\cite{Lerose2021, Giudice2022, PhysRevX.13.041008}. Understanding how these scaling behaviors arise due to the dynamics underlying a quantum stochastic process may have implications for the numerical simulation of chaotic systems.

Distinct signatures of chaos can be seen from the properties of the PPE, provided that deterministic interventions are considered. We also compared the dependence of the PPE on deterministic and non-deterministic interventions, observing that the latter resulted in reduced mean and larger variance of the bipartite entanglement entropy. Monitored many-body systems---time-evolving systems with projective measurements acting at multiple times throughout the evolution---have been studied recently in the context of measurement-induced phase transitions~\cite{Skinner2019, PhysRevB.100.134306, PhysRevB.99.224307,PhysRevB.98.205136}. In these works, it has been shown that the action of measurements has a disentangling effect which competes with entanglement generation from unitary dynamics. A phase transition is known to occur when there is a sufficiently large measurement rate, resulting in area-law entanglement growth. This phase transition is strongly associated with information scrambling~\cite{PhysRevLett.125.030505}: for instance, it has been shown that non-interacting free fermion models have area-law entanglement growth for arbitrarily weak measurements~\cite{10.21468/SciPostPhys.7.2.024} indicating a potential method to distinguish chaotic and non-chaotic dynamics. The PPE we introduce here provides a general framework to analyze the conditional dynamics of monitored many-body states, incorporating arbitrary interventions such as stochastically applied measurements, and thus it may prove useful for future studies on the role of chaos in measurement-induced phase transitions.

We generated the PPE using a fixed basis of local operators which act on the butterfly space. Other approaches to generating ensembles may be considered; for instance, in Ref.~\cite{Ippoliti2022solvablemodelofdeep}, they considered the ensemble of states on the butterfly space after acting a fixed basis of projective operators on the output states of the process. Comparing the signatures of chaos from this ensemble to the PPE may be interesting and is left for future work. The use of analytically tractable toy models in Ref.~\cite{Ippoliti2022solvablemodelofdeep} may also be useful for providing more accurate benchmarks of chaotic behavior, as opposed to the Haar ensemble used here. We have also considered a single type of initial state throughout, the N\'eel state. In Appendix \ref{app:initial_state_dep}, we explore two other types of initial states which have difficulty distinguishing chaotic from non-chaotic behaviour: typical (Haar-random) states and ground states. This emphasises the need for weakly entangled, high energy initial states for these probes to distinguish chaotic dynamics from integrable dynamics. These conditions are typically satisfied, for example, by the product states which constitute standard initial conditions for quantum computation and simulation.

Finally, we comment on the experimental accessibility of the quantities discussed in this paper. The construction of the process tensor requires post-selection---multiple samples of the output state $|\ups_{R|\vec{x}}\rangle$ are required for each multi-time intervention $\vec{x}$ to compute averages over the ensemble. To perform these calculations on a quantum circuit would require a number of runs which is exponential in the total number of interventions $n_B$. Despite this difficulty, post-selection has been performed in a superconducting qubit processor~\cite{Koh_2023}; however, this is restricted to small circuit sizes. Recent studies on MITPs have used the linear cross-entropy~\cite{aaronson2016, PhysRevLett.130.220404} as an order parameter for probing volume- and area-law entanglement. This approach replaces an exponentially long quantum simulation with a classical simulation which, in general, also scales exponentially but can be performed more efficiently for Clifford circuits as has been demonstrated experimentally in Ref.~\cite{ kamakari2024}. Developing a similar order parameter to estimate higher-order moments of the PPE would be interesting. Crucially, our finite-size scaling of the PPE entanglement variance in Fig.~\ref{fig:PPE_t_L} suggests that differences between chaotic and non-chaotic dynamics become exponentially large in system size, even for a relatively modest number of interventions, $n_B\sim 10$. This may prove beneficial for experimental observation of our predictions.

\begin{acknowledgments}
We are grateful to Philipp Strasberg and Dominik Safranek for discussions and collaboration on related topics. POD is supported by the Irish Research Council (ID GOIPG/2023/3847). ND acknowledges support of an Australian Government Research
Training Program Scholarship and the Monash Graduate Excellence Scholarship. ND further acknowledges funding by the Deutsche Forschungsgemeinschaft (DFG, German Research Foundation) under Germany’s Excellence Strategy - Cluster of Excellence Matter and Light for Quantum Computing (ML4Q) EXC 2004/1 - 390534769. KM is grateful for the support of the Australian Research Council Discovery Projects DP210100597. MTM is supported by a Royal Society-Research Ireland University Research Fellowship. We acknowledge the Irish Centre for High End Computing and Kesha for the use of their computational facilities.
\end{acknowledgments}

\section*{Data Availability}
The data that support the findings of this article are
openly available \cite{ODonovan2026code}.

\appendix
\renewcommand{\thesubsection}{\Roman{subsection}}

\section{Comparison of QDE Definitions}
\label{Appendix:ALF_vs_PT}

As discussed in the main text, similar definitions of the dynamical entropy have been studied previously for different systems and choices of interventions. In this section, we discuss the differences and similarities between the Alicki-Fannes dynamical entropy~\cite{lindblad_dyn_ent, Alicki1994-dc} and QDE defined from the pure state process tensor. For a summary, see Table~\ref{tab:AF_PT_comparison}.

The Alicki-Fannes dynamical entropy is defined as the supremum of the the Von Neumann entropy of the correlation matrix over the set of interventions which forms a partition of unity---a set of operators $\mathbf{A} = \{A_{1},...,A_{n}\}$ such that $\sum_{i} A^{\dagger}_{i}A_{i} = \mathds{1}$. The correlation matrix with respect to these operators has matrix elements given by the following,
\begin{equation}
    C\left[\mathbf{A}\right]_{\vec{i} \; \vec{j}} = \text{tr}\left[\rho A^{\dagger}_{i_1}(t_1)...A^{\dagger}_{i_n}(t_{n}) A_{j_n}(t_n)...A_{j_{1}}(t_1)\right].
\end{equation}
Above, the dynamics is written in the Heisenberg picture such that $A_{i_{k}}(t_{k}) = e^{iHt_{k}}A_{i_{k}}e^{-iHt_{k}}$ and $\rho$ is the state which is taken to be in equilibrium. The dynamical entropy is then defined as
\begin{equation}
\label{eq:ALF}
    S_{\text{dyn}} = \text{sup}_{\mathbf{X}}\limsup_{n \rightarrow \infty} \frac{1}{n}S(C\left[\mathbf{X}\right]),
\end{equation}
with $S(C\left[\mathbf{X}\right]) = -\text{tr}(C\left[\mathbf{X}\right] \log(C\left[\mathbf{X}\right]))$ the Von Neumann entropy taken with respect to the correlation matrix which is normalized to have unit trace.

A desirable property of the Alicki-Fannes entropy is that it has been shown to be equivalent to the classical dynamical entropy in the classical limit~\cite{Pechukas1982, Alicki1994-dc, Slomczynski1994-ns}. This contrasts the definition of the QDE used in this paper, which utilizes the R\'enyi-2 entropy to quantify entanglement in time. In general, this quantity will not become equivalent to the classical dynamical entropy as this is uniquely defined by the Shannon entropy. Additionally, the Alicki-Fannes entropy is defined for equilibrium states which is relevant in the classical limit as classical phase space distributions should be time-invariant. In this work, we consider pure initial states which are not time-invariant in general.

In theorem 10.2 of Ref.~\cite{q_dyn_sys_Alicki}, it is shows that the Alicki-Fannes entropy is zero in finite systems, independent of the type of dynamics chosen. This is due to an upper bounds on the entropy of the correlation matrix in finite-sized systems,
\begin{equation}
    S(C[\mathbf{X}]) \leq \text{min}\{\log(d_{R}), \log(d_{B})\}.
\end{equation}
Inserting this result into Eq.~\eqref{eq:ALF}, we see that 
\begin{equation}
    S_{\text{dyn}} = \limsup_{n\rightarrow \infty} \frac{\log(d_{R})}{n} \rightarrow 0.
\end{equation} 
When analyzing the dynamical entropy in finite-size systems, the condition that $n \rightarrow \infty$ is relaxed.

In Eq.~\eqref{eq:ALF}, the partition is comprised of operators which may act on the entire remainder space. In practice, it is numerically difficult to compute the dynamical entropy for arbitrary non-local interventions. Additionally, it is assumed that the supremum is taken over the possible partition $\mathbf{X}$ which is numerically challenging. In this work, we consider a restricted form of the dynamical entropy where partition of unitary is defined over a set of physically accessible interventions that are orthogonal and act on a local subspace $\mathcal{H}_{S}$ of the total remainder space $H_{R} \equiv \mathcal{H}_{SE}$. This restricted dynamical entropy has been studied in free fermion systems and has been computed in the thermodynamic limit \cite{Cotler2018}.

\begin{table}[h]
    \centering
    \begin{tabular}{|c||c|}
        \hline
         $\textbf{Alicki-Fannes}$ & \textbf{Process Tensor} \\
        \hline
        \hline
         Equilibrium & Arbitrary  \\
         \hline 
         Supremum & No Supremum\\
         \hline
        $ A_{i_{k}} \in \mathcal{L}(\mathcal{H}_{SE})$ & $A_{i_{k}} \in \mathcal{L}(\mathcal{H}_{S})$ \\
         \hline
    \end{tabular}
    \caption{Comparison of the definitions of the Alicki-Fannes dynamical entropy and the process tensor dynamical entropy which we use throughout this work, see Sec.~\ref{sec:DE} for more information.}
    \label{tab:AF_PT_comparison}
\end{table}

\section{Equilibration of Dynamical Entropy with time}

\label{Appendix:dyn_growth_t}
In this section, we derive the equilibration results discussed in Sec.~\ref{sec:results}. Equilibration results have been applied to the observable entropy in the single-time case~\cite{PRXQuantum.6.010309, Linden2008}. Here we use this approach combined with results from the theory of equilibration of process tensors \cite{Dowling2023relaxationof,finitetime} to show that the dynamical entropy will, for most times, be near its equilibrium entropy.

First, we use continuity bounds to relate the difference between the entropy of two states to the distance between the states. For R\'enyi-$\alpha$ entropy's with $\alpha > 1$, the following continuity bound has been derived~\cite{Chen_2016},
\begin{equation}
\begin{aligned}
    &|S^{(\alpha)}(\rho) - S^{(\alpha)}(\sigma)| \\
    &\leq \frac{d^{\alpha -1}}{1 - \alpha}\left[1- (1 -D)^{\alpha} - (d-1)^{1-\alpha} D^{\alpha} \right].
\end{aligned}
\end{equation}
Above, $d$ is the dimension of the Hilbert space, and $D$ is the trace norm distance between $\rho$ and $\sigma$ give by
\begin{equation}
    D = \frac{\|\rho - \sigma\|_1}{2} = \frac{1}{2}\tr\left[\sqrt{(\rho - \sigma)(\rho - \sigma)^{\dagger}}\right].
\end{equation} 
 In this work, we consider the $\alpha = 2$ case which gives the following bound,
\begin{equation}
\label{Appendix:eq:renyi_bound}
    |S^{(2)}(\rho) - S^{(2)}(\sigma)| \leq 2D - \frac{d-2}{d - 1} D^2.
\end{equation}

We will now show that the equilibration of the dynamical entropy occurs on average. To do this we consider the equilibrium process tensor defined as the infinite-time averaged $n_B$-process tensor \cite{Dowling2023relaxationof},
\begin{equation}
    \Omega_{B} = \langle \ups\rangle_{\infty}= \left(\prod_{i=1}^{n_B} \lim _{T_i \rightarrow \infty} \frac{1}{T_i} \int_0^{T_i} d\left(\delta t_i\right)\right) \Upsilon_{B}.
\end{equation}
The entropy of the equilibrium process tensor will correspond to the equilibrium entropy which the dynamical entropy will fluctuate around. We use the continuity bound in Eq.~\eqref{Appendix:eq:renyi_bound} to write the time averaged difference between the entropy of the process and the equilibrium process in terms of the difference between the processes. First we show that the right hand side of Eq.~\eqref{Appendix:eq:renyi_bound} is a concave function of the trace distance $T$. Taking the second derivative with respect to $T$ and find
\begin{equation}
\begin{aligned}
    f(T) = 2 D - \frac{d-2}{d-1}D^2 \rightarrow f''(D) = -2 \frac{d-2}{d-1} < 0,
\end{aligned}
\end{equation}
provided that $d \geq 2$. Using this result, Jensen's inequality ensures $\langle f(D)\rangle_{\infty} \leq f(\langle D\rangle_{\infty})$ such that the following inequality is satisfied
\begin{equation}
    \langle |S^{(2)}(\ups_{B}) - S^{(2)}(\Omega_{B})|\rangle_{\infty} \leq   2\langle D\rangle_{\infty} - \frac{d_{S}^{k}-2}{d_{S}^{k}-1} \langle D\rangle_{\infty}^2.
\end{equation}
We have written the time averaged distance between the entropy's as the time averaged distance between states, $\langle D\rangle_{\infty} = \frac{1}{2}\langle ||\ups_{k} - \Omega_{k}||_{1}\rangle_{\infty}$.

To apply the results of Ref.~\cite{Dowling2023relaxationof}, we first show that the trace distance can be rewritten in terms of the diamond norm distance. The diamond norm distance is the distance between two processes.
\begin{equation}
\label{Appendix:eq:diamond_norm}
    D(\ups_{B}, \Omega_{B}) = \frac{1}{2}\text{max}_{A_{k}} \sum_{\vec{x}} |\tr\left[A_{\vec{x}} (\ups_B - \Omega_B)\right]|.
\end{equation}
Given a set of POVM's $\{\Pi_{i}\}$, the following inequality holds,
\begin{equation}
    \frac{1}{2}||\ups_B - \Omega_{B}||_1 \geq \frac{1}{2} \sum_{i} |\text{tr}(\Pi_i (\ups_B - \Omega_{B}))|.
\end{equation}
When $\Pi_{i}$ is a projector onto the eigenvalues of $\Upsilon_{k} - \Omega_{k}$, then equality holds~\cite{Wilde_2013}. The maximum over instruments is being taken in Eq.~\eqref{Appendix:eq:diamond_norm}, so the diamond norm and trace norm are equivalent. We can now write the average difference of the entropy's in terms of the average diamond norm. 

Finally, we can use results obtained from the equilibration of process tensors to show that this difference is small on average. The following result was proven in \cite{Dowling2023relaxationof},
\begin{equation}
\label{eq:appendix:distance_bound}
    \langle D_{M_k}(\ups_B, \Omega_B)\rangle_{\infty} \leq \frac{1}{2}M_kd^{k}_S \sqrt{\frac{2^k - 1}{d_{\text{eff}}(\rho)}}.
\end{equation}
Here, $M_{k}$ is the combined total number of measurement outcomes in $M_{k}$. The effective dimension of the initial state is a measure of its overlap with the eigenstates of the Hamiltonian. The overlap of the state and the eigenstate is given by  $\$(\rho) = \sum_{n}P_{n} \rho P_{n}$ for $P_{n} = |n\rangle \langle n|$ the projector onto an eigenstate of the Hamiltonian. The effective dimension is the defined as 
\begin{equation}
    d_{\text{eff}}(\rho) = \frac{1}{\tr\left[\$(\rho)^2 \right]}.
\end{equation}

Given this bound, we can apply Markov's inequality. Consider a random variable $X$ with mean, $\mu$, and standard deviation, $\sigma$. Markov inequality is a bound on the probability of the random variable:
\begin{equation}
    \label{app:eq:markov}
    \mathds{P}(X \geq a) \leq \frac{E[X]}{a}.
\end{equation}
We can straightforwardly apply this inequality to the random variable $X = |S(\ups_{B}) - S(\Omega_B) |$. Here, we take 
\begin{equation}
    a = \frac{\langle |S^{(2)}(\ups_B) - S^{(2)}(\Omega_B)|\rangle_{\infty}}{\langle D\rangle_{\infty}\sqrt{d_{\text{eff}}}} = \frac{1}{d_{\text{eff}}^{1/4}}\left(2-\frac{d_{S}^{n_B}-2}{d_{S}^{n_B}-1}\langle D \rangle_{\infty} \right),
\end{equation}
which gives us Eq.~\eqref{eq:markov_bound} from the main text.

\section{Dynamical Entropy Rate as a Quantifier of Non-Markovianity}
\label{Appendix:Dyn_Ent_markov}

In this section we show that the rate of growth of the dynamical entropy places a bound on Non-Markovian correlations. First, consider a process $\Upsilon_{RB}$ with $d_{R} > d_{B}$ and subspaces $B_{1}$ and $B_{2}$ such that $t_{B_1} < t_{B_2}$. The quantum mutual information quantifies an unambigous, operational notion of non-Markovianity  in open quantum systems~\cite{processtensor2}, and it is given by
\begin{equation}
    I(B_1:B_2) = S_{\text{vN}}(\ups_{B_1}) + S_{\text{vN}}(\ups_{B_2}) - S_{\text{vN}}(\ups_{B}).
\end{equation}
Here, $S_{\text{vN}}(\ups_{B}) = E(B:R)$ is the dynamical entropy, $S_{\text{vN}}(\ups_{B_1})$ and $S_{\text{vN}}(\ups_{B_2})$ are the entropy of marginals of the process tensor.

Using causality conditions, we know that $S_{\text{vN}}(\ups_{B_1})$ is the QDE for a process with $n_{B_1}$ interventions. Using this fact, we rewrite the mutual information as follows,
\begin{equation}
    I(B_1:B_2) = S_{\text{vN}}(\ups_{B_2}) - n_{B_2} \frac{S_{\text{vN}}(\ups_{B}) - S_{\text{vN}}(\ups_{B_{1}})}{{n_{B} - n_{B_1}}}.
\end{equation}
We define the dynamical entropy growth rate as
\begin{equation}
    \mathcal{D}_{B_1:B_2}S_{\text{vN}}(\ups) = \frac{S_{\text{vN}}(\ups_{B})- S_{\text{vN}}(\ups_{B_1})}{n_{B} - n_{B_1}}.
\end{equation}
Finally, using that entropy is bounded from above by $S_{\text{vN}}(\ups_{B_2}) \leq \log(d_{B_{2}}) = n_{B_2} \log(d_{S})$,
we get the following expression for the mutual information
\begin{equation}
\label{Appendix:eq:NM_bound}
    I(B_1:B_2)\leq n_{B_2}\bigl[\log(d_{S}) - \mathcal{D}_{B_1:B_2}S_{\text{vN}}(\ups)\bigr].
\end{equation}

The mutual information is non-negative due to subadditivity, therefore the inequality here must be saturated when the right-hand side of Eq.~\eqref{Appendix:eq:NM_bound} is zero which corresponds to having a maximal growth rate,
\begin{equation}
    \mathcal{D}_{B_1:B_2} S_{\text{vN}}(\ups) = \log(d_{S}).
\end{equation}
We can conclude that maximal growth rate of dynamical entropy is a sufficient, but not necessary, condition for Markovianity. We have seen that chaotic systems maximize the dynamical entropy growth rate after sufficiently long time scales, indicating that chaotic models are Markovian.

\section{Spatiotemporal Entanglement for Haar Averaged Dynamics}
\label{Appendix:st_ent_haar}

In this section, we compute an estimate of the Haar averaged entropy of the reduced process $\ups_{R_1}$. The pure state process tensor is given by the following
\begin{equation}
    |\ups\rangle = \sum_{\vec{x}}UA_{x_n}U...UA_{x_1}U|\psi\rangle \otimes |x_n...x_1\rangle.
\end{equation}
The operators $A_{x_j}$ obey the following properties $\sum_{x_{j}} A^{\dagger}_{x_j} A_{x_j} = \mathds{1}_{R}$ and $\tr(A^{\dagger}_{x_{j}} A_{y_{j}}) = \delta_{x_{j} y_{j}}$.
We perform a Haar average over the unitary evolution and will assume that the unitary at each time step is different for simplicity. The Haar average of this process is given by
\begin{equation}
\begin{aligned}
     &\langle \ups \rangle_{\text{Haar}}= \sum_{\vec{x}, \vec{y}} \int_{U_{1:n+1}} U_{k+1}A_{x_n}...A_{x_1}U_{1} \Psi U_{1}^{\dagger} A_{y_1}^{\dagger}... A_{y_n}^{\dagger} U_{n+1}^{\dagger} \\
    & \qquad \qquad\qquad\qquad \otimes |x_{1}...x_{n}\rangle \langle y_{1} ... y_{n}|
\end{aligned}
\end{equation}
with initial state $\Psi = |\psi\rangle \langle \psi|$ and Haar average taken over each unitary independently. Starting from the middle, we compute the Haar integral over $U_{1}$ which is given by the following,
\begin{equation}
    \int_{U_1} U_1|\psi\rangle \langle \psi|U_{1}^{\dagger} = \frac{1}{d_{R}} \langle \psi|\psi\rangle \mathds{1} = \frac{1}{d_{R}} \mathds{1}.
\end{equation}
Here, $d_R$ the dimension of the remainder space. Inserting this into the integral we get
\begin{equation}
\begin{aligned}
    &\langle \ups \rangle_{\text{Haar}} = \frac{1}{d_{R}} \sum_{x, y}\int_{U_{2:n+1}} U_{n+1}A_{x_n}...U_{2}A_{x_1} \mathds{1} A_{y_1}^{\dagger} U_{2}^{\dagger}... A_{x_n}^{\dagger} U_{n+1}^{\dagger} 
    \\& \qquad \qquad \qquad \qquad \qquad  \otimes |x_{1}...x_{n}\rangle \langle y_{1} ... y_{n}|.
\end{aligned}
\end{equation}
Taking the average over $U_{2}$, we get
\begin{equation}
\begin{aligned}
    &\int_{U_2} U_{2}  \left[\mathds{1}_{E} \otimes A_{x_1}A^{\dagger}_{y_1} \right] U_{2} \\ 
    &= \frac{1}{d_R}\mathds{1}_{R} \tr_{E}(\mathds{1}_{E}) \tr_{S}(A_{x_1} A^{\dagger}_{y_1}) = \frac{d_{E}}{d_{R}}  \delta_{x_{1} y_{1}}\mathds{1}_{R}.
\end{aligned}
\end{equation}
Here, $d_{E} = d_{R}/d_{S}$ and is the dimension of the environment. We can repeatedly apply this result for all unitaries so that the average process is given by
\begin{equation}
\label{Haar_PT}
\begin{aligned}
    &\langle \ups \rangle_{\text{Haar}} = \left[\frac{d_E}{d_R}\right]^{n} \left[\frac{1}{d_R}\right] \mathds{1}_{R}\otimes \sum_{x,y} \delta_{\vec{x} \vec{y}} |\vec{x}\rangle \langle \vec{y}| \\
    & \qquad \quad \; \; = \frac{1}{d_R} \mathds{1}_{R}\otimes \left[\frac{1}{d_S}\right]^n\mathds{1}_{B}.
\end{aligned}
\end{equation}
The expression in Eq.~\eqref{Haar_PT} is the Haar average process tensor.

Using Eq.~\eqref{Haar_PT}, we can estimate different entropic quantities of the process tensor. In particular, we can estimate $S(R_1)$ for which we obtain
\begin{equation}
    \langle S^{(2)}(\ups_{R_1})\rangle \approx -\log(\tr\left[\langle \ups_{R_1}\rangle^2\right]) = \log(d_{R_1}).
\end{equation}
This result is used in Sec.~\ref{sec:STE} to compare with our numerical results.

\section{Moments of Projected Process Ensemble}
\label{Appendix:PPE_moments}

In the previous section, we have computed the Haar averaged pure process tensor. This corresponds to the first moment of the PPE, $\mathcal{E}$. In this section, we will compute compute higher order moments of the PPE and use them as a probe of chaos. We first compute the second moment of the PPE which is given by the following
\begin{equation}
    \ups^{(k)}_{R} = \sum_{\vec{x}} p_{\vec{x}} |\ups_{R|\vec{x}}\rangle \langle \ups_{R|\vec{x}}|^{\otimes k}.
\end{equation}
The state $|\Upsilon_{R|\vec{x}}\rangle$ is constructed by projecting onto a basis of input interventions $|\vec{x}\rangle$ on butterfly space and is normalized. We will assume that the interventions are unitary such that $p_{\vec{x}} = \frac{1}{d_S^n}$ will average over the unitary operators individually as shown below
\begin{equation}
\begin{aligned}
    &\langle \ups^{(k)}_{R}\rangle_{\text{Haar}} = \frac{1}{d_S^n}\sum_{\vec{x}} \int_{U_{1:n+1}}  \left[U_{n+1}A_{x_{n}}... A_{x_1}U_{1} |\psi\rangle\right]^{\otimes k} 
    \\& \qquad \qquad \qquad \qquad \quad  \times  \left[\langle \psi|U_{1}^{\dagger} A_{x_1}^{\dagger} ... A_{x_{n}}^{\dagger} U^{\dagger}_{n+1}\right]^{\otimes k}
\end{aligned}
\end{equation}
For simplicity, we write the integrand tensor product as
\begin{equation}
    \ups^{(k)}_{R} = U_{n+1}^{\otimes k}  ...A_{x_{1}}^{\otimes k} U_{1}^{\otimes k} \Psi^{\otimes k} (U_{1}^{\dagger})^{\otimes k} (A^{\dagger}_{x_{1}})^{\otimes k}... (U_{n+1}^{\dagger})^{\otimes k}
\end{equation}

\begin{figure*}[t]
    \centering
    \includegraphics[width=0.95\textwidth]{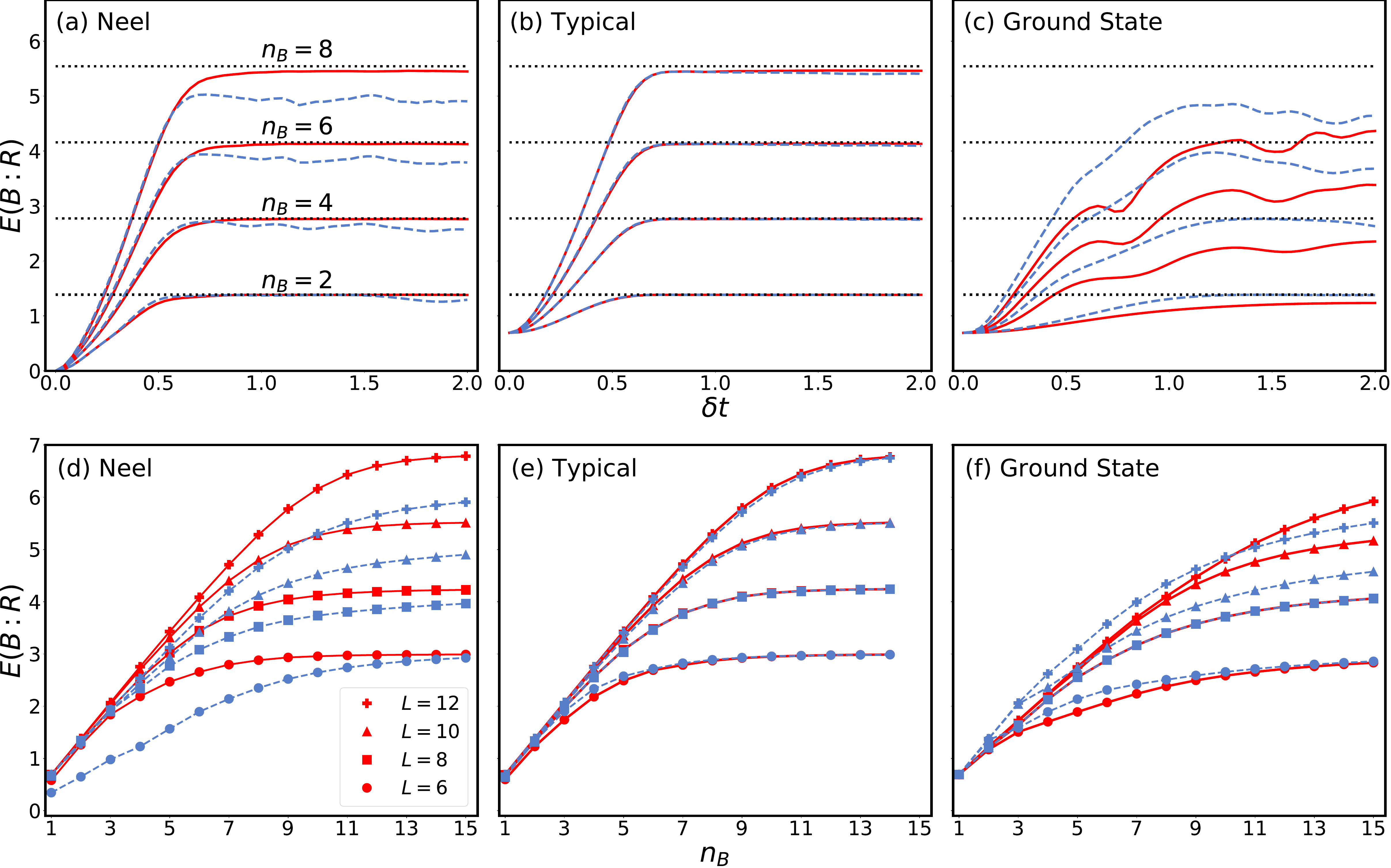}
    \caption{QDE computed using N\'eel (a, d), typical (b,e), and ground (c, f) states. We consider the integrable and chaotic XXZ model for these simulations which are shown in blue and red respectively. We plot QDE for as a function of time between interventions in the top row for system size $L = 14$. We observe that the typical state preparation results in maximal growth for both the chaotic and integrable models. The ground state preparation displays distinctly less QDE growth with time with even the integrable model having larger growth. The scaling with number of interventions shown in the bottom row, we see again that the typical state case shows linear behaviour for integrable and chaotic models and that the ground state case shows reduced growth rate for the chaotic model.}
    \label{fig:dyn_ent_state_dep}
\end{figure*}

Methods from the group theory of symmetric subspaces~\cite{Church_of_sym} can be used to simplify the calculation of this Haar averaging. Consider again the middle integral over $U_1$, we can write this averaging as a map,
\begin{equation}
    \Phi_1[\Psi^{\otimes k}] = \int_{U_1}(U_1)^{\otimes k} \Psi^{\otimes k} (U_1^{\dagger})^{\otimes k},
\end{equation}
 with $\Phi_1: \mathcal{L}\left[\mathcal{H}_{R}^{\otimes k}\right] \rightarrow \mathcal{L}\left[\mathcal{H}_{R}^{\otimes k}\right]$ mapping from the space of operators 
 acting on $H^{\otimes k}_{R}$ to itself. The Haar average is invariant under all unitary transformations $V^{\otimes k}$ such that
\begin{equation}
    \left[\Phi_1[\Psi^{\otimes k}], V^{\otimes k} \right] = 0
\end{equation}
for all $V^{\otimes k} \in \mathcal{U}^{\otimes k}$. Schur's lemma then implies that the Haar average of this state must be proportional to the identity over this space. It is known that the unitary group $\mathcal{U}$ under the action $\mathcal{U} \rightarrow \mathcal{U}^{\otimes k}$ has the symmetric subspace of $\mathcal{H}^{\otimes k}_{R}$ as an irreducible representation. The identity on this subspace is given by
\begin{equation}
    \Phi_{1}[\Psi^{\otimes k}] \propto \mathds{1}_{S_k} = \sum_{\pi \in S_{k}}P_{\pi},
\end{equation}
where $P_{\pi}$ is a representation of the symmetry group $S_{k}$ which acts as
\begin{equation}
    P_{\pi}|\psi_1 \rangle \otimes...\otimes |\psi_k\rangle = |\psi_{\pi^{-1}(1)}\rangle \otimes ...\otimes |\psi_{\pi^{-1}(k)}\rangle.
\end{equation}
We note that this calculation has been performed previously in the context of projected ensembles \cite{Cotler2018}.

We can use the property $\left[A_{x_{1}}^{\otimes k}, P_{\pi} \right] = 0$ to write the action of the CP maps as
\begin{equation}
\begin{aligned}
    &\sum_{x_{1}} A_{x_{1}}^{\otimes k} \Phi_{1}[\Psi^{\otimes k}] (A^{\dagger}_{x_{1}})^{\otimes k} = \Phi_{1}[\Psi^{\otimes k}]\sum_{x_{1}}(A_{x_{1}}A^{\dagger}_{x_{1}})^{\otimes k}.
\end{aligned}
\end{equation}
The next integral can also be represented straightforwardly as 
\begin{equation}
\begin{aligned}
    \int_{U_2} (U_2)^{\otimes k} \left[\sum_{x_{1}} A_{x_{1}}^{\otimes k} \Phi_{1}[\Psi^{\otimes k}] (A^{\dagger}_{x_{1}})^{\otimes k}\right](U_2^{\dagger})^{\otimes k} 
    &\\= \Phi_{1}[\Psi^{\otimes k}]\sum_{x_{1}}\int_{U_2} (U_{2})^{\otimes k} (A_{x_{1}} A^{\dagger}_{x_{1}})^{\otimes k} (U_{2}^{\dagger})^{\otimes k}
\end{aligned}
\end{equation}
This Haar average is given by the identity over the symmetric subspace and is represented as $\Phi_2[(A_{x_1}A_{x_1}^{\dagger})^{\otimes k}]$. We note that this result is independent of $x_1$ so we can write
\begin{equation}
    \sum_{x_1} \Phi_2[(A_{x_1}A_{x_1}^{\dagger})^{\otimes k}] = d_S \Phi_2[(A_{x_1}A_{x_1}^{\dagger})^{\otimes k}]
\end{equation}

Repeatedly applying this result, we get the following expression for the $k$-th moment of the PPE,
\begin{equation}
    \langle \ups^{(k)}_{R}\rangle_{\text{Haar}} = d_{S}^n \Phi_{1}(\Psi^{\otimes k}) \prod^{n}_{i = 1}\Phi_{i+1}[(A_{x_{i}}A_{x_i}^{\dagger})^{\otimes k}].
\end{equation}
Using the fact that $\sum_{\pi, \pi'} P_{\pi} P_{\pi'} \propto \sum_{\pi} P_{\pi}$, we can show that the Haar averaged $k$-th moment is proportional to the following,
\begin{equation}
    \langle \ups^{(k)}_{R}\rangle_{\text{Haar}} \propto \sum_{\pi \in S_{k}} P_{\pi}.
\end{equation}
The $k$-th order moment should have unit trace and using properties of the symmetric group we find the normalization to be given by the following expression
\begin{equation}
\label{appendix:eq:haar_kth_moment}
    \langle \ups^{(k)}_{R}\rangle_{\text{Haar}} = \frac{\sum_{\pi \in S_{k}} P_{\pi}}{(d_R)(d_R + 1)...(d_R +k-1)}.
\end{equation}

\begin{figure*}[t]
    \centering
    \includegraphics[width=0.95\textwidth]{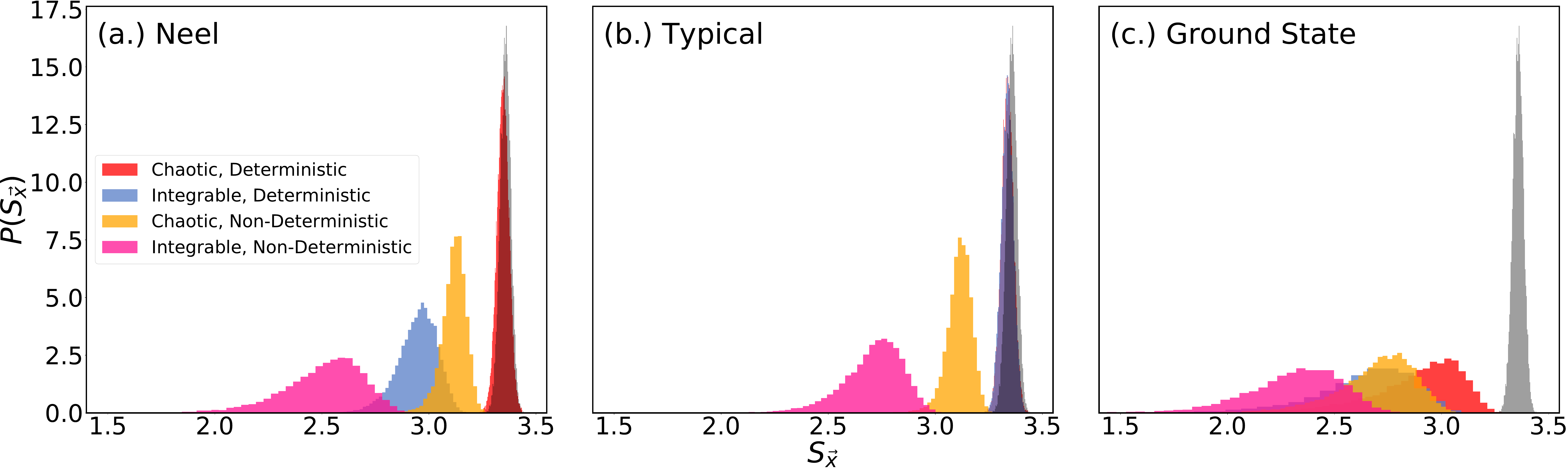}
    \caption{Bipartite entanglement distribution computed from the PPE using using N\'eel (a), typical (b), and Ground (c) state preparations.  We consider chaotic and integrable XXZ models with system size $L = 12$ and number of interventions $n_{B} = 14$. (a.) We observe the same features as seen in Sec.~\ref{results:HOM}, agreement between the Haar distribution and chaotic dynamics for deterministic interventions. (b.) For the typical initial state, we see the deterministic interventions show behaviour similar to the Haar ensemble, independent of the chaoticity of the dynamics. For non-deterministic interventions, we see large we can observe the competition between entanglement breaking projective measurements and entanglement generating dynamics, from which we can distinguish chaos from integrability. Finally, for the ground state (c.) we observe that all models show distinct behaviour to the Haar ensemble, but we are still able to distinguish chaotic and integrable dynamics by proximity to the Haar ensemble.}
    \label{fig:PPE_dist_state_dep}
\end{figure*}

\section{Initial State Dependence}
\label{app:initial_state_dep}

In this section, we discuss the dependence of the QDE and PPE on the choice of initial state. So far, we have only used the N\'eel state as our initial preparation. This state has low initial entanglement and an energy which is near the center of the spectrum. These are favourable properties for distinguishing chaos from integrability. Initial states with a large initial entanglement can often exhibit chaotic behaviour, even for integrable dynamics, as they have intrinsic complexity which is difficult to distinguish from the complexity of the dynamics. Additionally, initial states near the edges of the spectrum may exhibit constrained dynamics, even for chaotic dynamics. We will now perform a numerical analysis to see if the quantities considered in this work are capable of distinguishing chaos from integrability under these conditions.

Below, we will compare our numerical results for the N\'eel state to the case with the two "bad" initial states, namely typical states and ground states. A typical state---one that is randomly sampled from the Hilbert space---has extensive initial entanglement and may exhibit chaotic behaviour for integrable dynamics. The ground state, by contrast, may exhibit highly constrained dynamics independent of the chaoticity of the model considered.

First, we study the QDE for the chaotic and integrable XXZ model using these different initial states. In Fig.~\ref{fig:dyn_ent_state_dep}, we reproduce the time dependence of the QDE and the growth with number of interventions for these initial states. We see that typical initial states, shown in Figs.~\hyperref[fig:dyn_ent_state_dep]{7b} and  \hyperref[fig:dyn_ent_state_dep]{7e}, exhibit indistinguishable behaviour for the two types of models. This is in stark contrast to Figs.~\hyperref[fig:dyn_ent_state_dep]{7a} and \hyperref[fig:dyn_ent_state_dep]{7d} which are the results for the N\'eel state. For the ground state preparation, we observe sub-maximal growth of the chaotic dynamical entropy and even see the integrable having larger growth rate with number of interventions. This clearly indicates that the QDE requires initially low entangled states with energy near the center of the spectrum to distinguish chaos and integrability in finite-sized systems.

We also study the PPE for these initial states. In Fig.~\ref{fig:PPE_dist_state_dep}, we plot the bipartite entanglement distributions for deterministic, $\{\mathds{1}_1, \sigma^z_1\}$, and non-deterministic, $\{|0\rangle \langle 0|, |1\rangle \langle 1|\}$, interventions. In Fig.~\hyperref[fig:PPE_dist_state_dep]{8b}, the non-deterministic interventions are able to distinguish chaotic and integrable dynamics, but the deterministic interventions show similar behaviour to the Haar ensemble, independent of the choice of dynamics. The ground state case in Fig.~\hyperref[fig:PPE_dist_state_dep]{8c} displays clear differences between the Haar ensemble and all models considered unlike what is seen for the N\'eel state.

In this analysis, we have found that the probes considered in this work require a low-entangled initial state with energy near the center of the spectrum. However, this requirement is not unique to the quantifiers considered in this work. If we assume that only chaotic dynamics can generate complexity, then choosing a volume-law entangled initial state is consistent with assuming that some chaotic dynamics was initially performed on the state which the probe is detecting. Similarly, chaos quantifiers like the eigenstate thermalization hypothesis also neglect states near the edges of the spectrum.

\bibliography{bib-file}

\end{document}